# A billion or more years of possible periglacial/glacial cycling in Protonilus Mensae, Mars

R.J. Soare,[1] J-P Williams,[2] A.J. Hepburn,[3] F.E.G. Butcher[4]

[1]Geography Department, Dawson College, Montreal, Qc, Canada, H3Z 1A4
(rsoare@dawsoncollege.qc.ca)

[2]Department of Earth, Planetary and Space Sciences, University of California,
Los Angeles, CA, USA 90095

[3]Department of Geography, Durham University, Durham, United Kingdom DH1 3LE

[4]Department of Geography, University of Sheffield, Sheffield, United Kingdom S10 2TN

**Pages:** 71

**Figures:** 15

**Abstract**

The long-term cyclicity and temporal succession of glacial-periglacial (or deglacial) periods or epochs are keynotes of Quaternary geology on Earth. Relatively recent work has begun to explore the histories of the mid- to higher-latitudinal terrain of Mars, especially in the northern hemisphere, for evidence of similar cyclicity and succession in the Mid to Late Amazonian Epoch.

Here, we carry on with this work by focusing on *Protonilus Mensae* [*PM*] (43-49$^0$ N, 37-59$^0$ E). More specifically, we discuss, describe and evaluate an area within *PM* that straddles a geological contact between two ancient units: *[HNt]*, a Noachian-Hesperian Epoch transition unit; and *[eHT]* an early Hesperian Epoch transition unit. Dark-toned terrain within the *eHt* unit (*HiRISE* image ESP_028457_2255) shows continuous coverage by structures akin to *c*lastically-*s*orted *c*ircles [*CSCs*]. The latter are observed in permafrost regions on Earth where the freeze-thaw cycling of surface and/or near-surface water is commonplace and cryoturbation is not exceptional.

The crater-size frequency distribution of the dark-toned terrain suggests a minimum age of ~100 Ma and a maximum age of ~1 Ga. The age estimates of the candidate *CSCs* fall within this dispersion. Geochronologically, this places the candidate *CSCs* amongst the oldest periglacial landforms identified on Mars so far.

Unit *HNt* is adjacent to unit *eHt* and shows surface material that is relatively light in tone. The coverage is topographically irregular and, at some locations, discontinuous. Amidst the light-toned surface, structures are observed that are akin to clastically *n*on-*s*orted *p*olygons [*NSPs*] and polygonised thermokarst-depressions on Earth. Terrestrial polygon/thermokarst assemblages occur in permafrost regions where the freeze thaw cycling of surface and/or near-surface water is commonplace and the permafrost is ice-rich. The crater-size frequency distribution of the light-toned terrain suggests a minimum age of ~10 Ma and a maximum age of ~100 Ma. The age estimates of the candidate ice-rich assemblages fall within this dispersion. Geochronologically, this places them well beyond the million-year ages associated with most of the other candidate ice-rich assemblages reported in the literature.

Stratigraphically intertwined with the two possible periglacial terrains are landforms and landscape features (observed or unobserved but modelled) that are indicative of relatively recent glaciation ( ~10 Ma - 100 Ma) and glaciation long past (≥~1 Ga) to decametres of depth: glacier-(cirque) like features; viscous-flow features, lobate-debris aprons; moraine-like ridges at the fore, sides and midst of the aprons; and, patches of irregularly shaped (and possibly volatile-depleted) small-sized ridge/trough assemblages. Collectively, this deeply-seated intertwining of glacial and periglacial cycles suggests that the Mid to Late Amazonian Epochs might be more Earth-like in their cold-climate geology than has been thought hitherto.

## 1. Introduction

At the mid- to relatively high-latitudes of Mars' northern plains, surface textures, landscape features, landforms and spatially-continuous landform assemblages reminiscent of current and/or relict periglacial terrain on Earth have been reported widely (e.g. Costard and Kargel, 1995; Mustard et a., 2001; Seibert and Kargel, 2001; Soare et al., 2008, 2014, 2017, 2018; Balme et al., 2009; Levy et al., 2009a, b, 2010; Ulrich et al., 2010; Gallagher et al., 2011; Hauber et al., 2011; Séjourné et al., 2011, 2012; Barrett et al., 2017, 2018; Johnsson et al., 2018; Gastineau et al., 2020). Almost invariably, the terrain is nested in smooth and sparsely cratered material that mutes and blankets or mantles the local topography. This *mantle(s)* is(are) metres to decametres thick and is/are thought to be: **1)** composed of ice, dust or a combination derived therefrom; **2)** relatively youthful, i.e. almost present day - ≤ ~10 Ma, although some age estimates are slightly higher than this; **3)** accumulated cyclically at the Martian surface by way of atmospheric precipitation; and, **4)** engendered by periodic variances in the spin-axis tilt and orbital eccentricity of Mars (e.g. Mustard et al., 2001; Head et al., 2003; Milliken et al., 2003; Laskar et al., 2004; Schon et al., 2009).

Surface textures, landscape features, landforms and spatially-continuous landform assemblages reminiscent of current and/or relict glacial-regions on Earth are observed at or near the Mars dichotomy and throughout the mid-latitudes of the northern plains (e.g. Kargel and Strom, 1992; Head et al., 2002, 2003; Forget et al., 2006; Dickson et al., 2008; Morgan et al., 2009; Baker et al., 2010, 2015; Souness and Hubbard, 2013; Hubbard et al., 2014; Sinha and Murty, 2015; Brough et al., 2016; Hepburn et al., 2020; Soare et al., 2021b, c). Their estimated ages shows greater variance (almost present day - ~1 Ga) than the candidate periglacial terrain referenced above (e.g. Head et al., 2003; Morgan et al., 2009; Baker et al. 2010; Souness and Hubbard, 2013; Hubbard et al., 2014; Sinha and Murty, 2015; Brough et al., 2016).

The long-term cyclicity and temporal succession of periglacial-glacial periods or epochs are keynotes of cold-climate geology on Earth, especially as it pertains to the Quaternary Period. Relatively recent work has begun to explore terrain at or close to the Mars dichotomy for geological/geomorphological evidence of similar cyclicity and succession (e.g. Dickson et al., 2008; Morgan et al., 2009; Baker et al., 2010; Head et al., 2010; Souness and Hubbard, 2013; Levy et al., 2014; Sinha and Murty, 2015; Hepburn et al., 2020).

Here, we carry on with this work by exploring an area within *P*rotonilus *M*ensae [*PM*] that lies to the east of the Lyot impact crater, to the north of the Moreux impact crater, and adjacent to the Mars crustal-dichotomy **(Fig. 1)**. The latter is a global geological-boundary that separates the ancient southern highlands (Late Noachian-Early Hesperian or Middle Noachian Epochs (McGill and Dimitriou, 1990 and Frey et al., 2002, respectively) from the relatively young northern-lowland plains (Early Amazonian Epoch) (Head et al., 2002; Tanaka et al., 2014).

The focus of our interest is a sub-region of *PM* that straddles a geological contact (45.06º N and 42.20º E) between two ancient units: *[HNt]*, a Noachian-Hesperian Epoch transition unit; and *[eHT]* an early Hesperian Epoch transition unit (Tanaka et al., 2014) **(Fig. 2)**. Here, complex cross-cutting relationships and relative stratigraphies (inferred from and supported by modelling), complimented by a suite of crater-size frequency distribution [*CSFD*] age estimates, point to possible glacial and deglacial (or periglacial) cycles having taken place as far back into the Amazonian Epoch as ~1 Ga, possibly even earlier than that.

## 2. Methods

The *H*igh *R*esolution *I*maging *S*cience *E*xperiment [*HiRISE*] image ESP_028457_2255 (from the *M*ars *R*econnaissance *O*rbiter [*MRO*], McEwen et al., 2007) and *C*ontext *C*amera [*CTX*] image F21_044083_2248_XI_44N317W, also from the *MRO*, Malin et al., 2007) frame our study region. Crater counts were conducted on the *HiRISE* image (25 cm pix$^{-1}$). The *CraterTools* plug-in for the *ESRI ArcGIS* was used to measure crater diameters (Kneissl et al., 2011). A 45 km$^2$ region of dark-toned and slightly elevated terrain north-northwest of the geomorphologic contact was identified for crater counts. The population of candidate craters with $D < 80$ m were divided into four classifications based on the presence or absence of morphologic characteristics diagnostic of an impact origin to rank the features from low to high confidence of an impact origin. Crater diameters were binned to generate cumulative and differential *c*rater *s*ize-*f*requency *d*istributions [*CSFDs*]. They were compared with modeled crater-retention age isochrons from Hartmann (2005) to provide estimates of crater retention ages.

With the aforementioned *CTX* image, we mapped the geomorphology in our study region in *ESRI ArcGIS Pro*. In total, seven units were mapped, distinguished according to systemic variations in surface texture visible at a 1:10,000 scale. The uncertainty in area associated with our mapping is less than a few percent, assuming a uniform 1-pixel (~5 m) misidentification along each boundary. For features with a curvilinear expression (e.g., supra-viscous-flow feature structure), individual landforms were digitized using a line along their length according to a perceived centerline, planform features were digitized using polygonal boundaries delineating their extent. *HiRISE* image ESP_028457_2255 was used to supplement our interpretation; however, we note that with only partial *HiRISE* coverage in our study region comprehensive mapping at the higher 25 cm pix$^{-1}$ resolution is not possible. Finally, elevation data was taken from the *H*igh-*R*esolution *S*tereo *C*amera [*HRSC*] (Neukum et al., 2004) Digital-elevation model [*DEM*] H1578_0000 (100 m pix$^{-1}$) referenced to the areoid. The vertical uncertainty associated with the *HRSC-DEM* is estimated to be 10 m. We compared all *HRSC* elevation measurements to the lower resolution (but more vertically accurate, ~3 m) *M*ars *O*rbiter *L*aser *A*ltimeter [*MOLA*] point data (Zuber et al., 1992).

The regional mapping of Tanaka et al. (2014) does not account for or comprise extant masses or bodies of icy materials at or near the surface of a geological unit let alone to depth. To estimate the reach of viscous flow-features possibly buried beneath the surface of unit *NHt* (**see section 9.2**) we used a 2D model of perfect plasticity calculate ice thickness on Earth (e.g. Ng and al., 2010; Benn and Hulton 2010) and Mars on (e.g. Parsons et al., 2011; Fastook et al., 2014; Karlsson et al., 2015; Schmidt et al., 2019; Hepburn et al., 2020a). The parabolas produced by these 2D approximations are good fits for contemporary lobate debris-apron topography. By inverting one such model for bed topography Karlsson et al., (2015) derived a mean yield stress for lobate debris aprons of $\tau y = 22$ kPa.

The model we use generates an estimated surface profile for a given glacier-reach informed by the mapping described above. We prescribe a driving yield stress of $\tau y = 22$ kPa, and assume the bed geometry is flat, a common assumption made when modelling ice masses on Earth (e.g. Hulton and Mineter, 2000; Cliffe and Morland, 2004). The model surface profile was then compared to the measured surface profile from the *MOLA* elevation data.

## 3. Observations

### 3.1 Unit eHt: surface structures and their morphologies

Adjacent to the western border of the geological contact separating units *eHt* from *HNt*, relatively dark-toned terrain is observed **(Fig. 2)**. The terrain is covered continuously by two principal landscape features:

1) Circular to sub-circular structures (~10-20 m in diameter), sometimes open/sometimes closed **(Fig. 3a)**; their distribution is continuous and limited to unit *eHT*. The structures have elevated margins or shoulders comprised of boulders (observed) and rock particles of lesser diameter (unobserved but deduced). Unobserved but deduced because the *HiRISE* camera cannot clearly resolve structures whose diameters are $<\sim91$ cm and, based on possible terrestrial analogues, it would be highly unusual for the margins not to comprise disparately-sized rock particles (e.g. Kleman and Borgström, 1990). The centre-fill material appears smooth, albeit at *HiRISE* resolution. As such, *smooth* fill could comprise rock-particle sizes anywhere below the near-metre scale of *HiRISE* resolution.

**2)** A second type of feature is more consistently circular and closed. It also has higher depth-to-width ratios, is bowl shaped and displays a greater variance of diameter than the first feature **(Figs. 3a-e)**. Some of these structures show inward-oriented terraces or benches and central mounds **(Figs. 4a-d)**. A possible sub-class of these structures (typically $D > 100$ m) comprise subdued, shallow circular depressions with fractures, and scarps **(Fig. 4d)**.

*3.2 Unit HNt: surface structures and their morphologies*

To the east of the geological contact separating units *eHt* and *HNt* lie multiple massifs covered discontinuously and surrounded (in an apron-like manner) by surface material that is relatively light in tone **(Figs. 2, 5a, d)**. The apron is demarcated at the fore, midst and sides by bouldered ridges that are roughly linear or curvilinear **(Figs. 5a-b, d)**. Upslope of the ridges and apron, and constrained within some massif valleys, possible flow lineations are observed **(Figs. 5a-b)**. Accumulations of snow, ice or debris cover that exhibit amphitheatre-like shape seem to head the candidate flow-lines near the massif summits **(Figs. 5a-b)**. Patchily distributed but spatially continuous assemblages of small-sized and geometrically-irregular ridges and troughs also occur throughout the basin **(Figs. 5a, c)**. Individual ridges and troughs are metres in elevation, metres to decametres in width and aggregated as closed or open structures **(Fig. 5e)**.

Polygonised and closed structures slightly smaller in diameter than those on the dark-toned terrain are nested patchily within the light-toned surface material **(Fig. 6)**. The polygons lack raised margins, let alone margins punctuated with boulders, and exhibit no apparent clastic sorting. However, some of the polygons are high-centred relative to their margins. At some locations, the polygons incise clustered and circular/sub-circular to elongate depressions that are metres deep **(Fig. 6)**.

## 4. Periglacial landscapes on Earth

*4.1 Clastically-sorted circles*

*C*lastically-*s*orted *c*ircles [*CSCs*] are a type of patterned ground uniquely associated with permafrost landscapes. Individual units, in the main, are ≤~10 m in diameter. *CSCs* are readily discerned by: **a)** the sharp contrast of rock particle sizes in the circle centres and margins; and, **b)** the positive elevation of the circle margins compared to the relatively-flat centres **(Fig. 7a)** (e.g. Ballantyne and Mathews, 1982, 1983; Washburn, 1989; Schlyter, 1992; Kruger, 1994). Typically, the centres comprise relatively fine-grained and frost susceptible particles with poor drainage potential (e.g. clays to silts to fine-sands); circle margins are elevated, relative to the centres, and are composed of rock particles or clasts that are larger than the centres (e.g. pebbles, cobbles or boulders) Ballantyne and Mathews, 1982, 1983; Washburn, 1989; Schlyter, 1992; Kruger, 1994). *CSC* distribution ranges from isolated, patchy or discontinuous to continuous and extensive, covering multiple square kilometres of terrain at some locations (e.g. Ballantyne and Mathews, 1982, 1983; Schlyter, 1992; Kruger, 1994).

The conditions or requirements needed for the origin and development of *CSCs* include (e.g. Ballantyne and Mathews, 1982, 1983; Kruger, 1994; Kling, 1996; Van Vliet-Lanoe, 1998):

**a)** relatively high soil moisture (at least intermittently);

**b)** iterative or episodic freeze-thaw cycling in the active layer of permafrost;

**c)** ice and soil segregation;

**d)** cryoturbation; and, possibly,

**e)** antecedent thermal-contraction (Kruger, 1994; Kling, 1996) or desiccation cracking (Ballantyne and Mathews, 1982, 1983). These processes facilitate the coalescence of cobbles or boulders into marginal patterns of distribution.

Interestingly, field observations in Scandinavia have shown that clastic sorting preceded coverage by Holocene-period glaciers at some locations (Whalley et al., 1981; Kling, 1996) and succeeded deglaciation at others (Ballantyne and Mathews, 1982, 1983; Kruger, 1994).

The mechanics of periglacially-constrained sorting are complex. One of the leading hypotheses is based on: **a)** water undergoing iterative freeze-thaw cycling; **b)** soil circulation and clastic up-freezing within the active layer of permafrost transporting larger sized clasts to the surface; and, **c)** radial displacement of the cobbles or boulders to the border or margins (Washburn, 1989; Pissart, 1990).

During top-down active-layer freezing, liquid water migrates towards the descending freezing-front and transient ice lenses form at various depths. As the descending freezing-front passes clasts, they are heaved upwards by the newly-formed ice lenses (e.g. Miller, 1972). During thaw, wet-fines flow and settle through clast interspaces. In subsequent episodes of freeze-thaw, clasts and fines are iteratively segregated, and clasts uplifted, by this ratcheting mechanism (e.g. Ballantyne and Harris, 1994). Also, as the freezing-front passes downwards through the active layer, size sorting may be achieved or aided by the different rates at which the freezing-front passes through clasts and wet, frost-susceptible fines. The pore-water surrounding wet fines must freeze before the freezing-front can pass through them but latent-heat transfers retard this process (e.g. Ballantyne and Harris, 1994). In clasts, the freezing-front propagates without this impediment and, consequently, moves more quickly than through a comparable volume of wet fines. As such, ice lenses can form and induce heave preferentially beneath clasts.

During subsequent episodes of freeze–thaw, and the iterative heaving and segregation of fines from clasts, vertical clasts collapse and creep horizontally outwards from uplifted centres of heave. This can lead to clast depletion over the heaving and slightly-elevated centres but clast concentration in a radially-expanding creeping front (e.g. Hallet et al., 1988). As neighbouring circles or polygons converge, clasts may be forced to build upward, forming raised and possibly imbricated clastic-borders (e.g. Dahl, 1966; Kessler and Werner, 2003).

## 4.2 Clastically non-sorted polygons

Clastically (non-sorted) polygons [*NSPs*], like *CSCs*, are ubiquitous features amidst permafrost landscapes on Earth (e.g. Lachenbruch, 1962; Czudek and Demek 1970; Washburn, 1973; Mackay, 1974; Rampton and Bouchard, 1975; Rampton, 1988; French, 2007) **(Fig. 7b)**. Generally ≤25 m in diameter, the polygons are produced by the tensile-induced fracturing of frozen sediment. This occurs when the latter undergoes a sharp drop of sub-zero (Celsius) temperatures (de Leffingwell, 1915; Lachenbruch, 1962). Fracturing, or *thermal-contraction cracking*, opens up shallow, narrow and vertical veins (Lachenbruch, 1962).

Iterative in-filling prevents the cracked ground from relaxing and returning to its initially seamless state as temperatures rise, diurnally or seasonally. As the iterative cycles increase in number, the shallow and narrow vertical veins may evolve into metres-wide and decametre-deep (vertically-foliated) wedges (e.g. Lachenbruch, 1962; Washburn, 1973; Mackay, 1974; French, 2007). Each of the foliations comprises the work of one fill cycle.

Fill-types vary. They are constrained by local or regional boundary conditions and by the availability of: **1)** meltwater derived of thawed snow or ice; vs, **2)** winter hoarfrost; vs, **3)** windblown sand, mineral-soil, or a mixture of the two (e.g. de Leffingwell, 1915; Péwé, 1959; Lachenbruch, 1962; Washburn, 1973; Sletten et al., 2003; Hallet et al., 2011).

As wedge-cracks become progressively dense in their distribution they intercept one another and form polygons (Lachenbruch, 1962). Some polygon networks are expansive, covering tens if not hundreds of km$^2$ in places like the Tuktoyaktuk Coastlands (e.g. Rampton , and Mackay, 1971; Mackay, 1974; Rampton, 1988) and are produced by countless iterations of seasonal or diurnal cracking and filling (e.g. Black, 1954; Lachenbruch, 1962; Washburn, 1973; Mackay, 1974).

Wedge growth, regardless of the fill type, is vertical and horizontal. As wedges aggrade at the polygon margins, their sedimentary overburden rise above the elevation datum of the polygon centres; this forms *low-c*entred *p*olygons [*LCPs*] (Péwé, 1959; Washburn, 1973; Harris et al., 1988; Rampton, 1988; French, 2007). Degradation, by thaw in the case of ice wedges or aeolian erosion in the case of sand or mineral wedges, depletes the wedge volume and mass and deflates the marginal overburden. *H*igh-*c*entred *p*olygons [*HCPs*] develop if and when this depletion and deflation lowers the polygon margins below the elevation of the centres (Péwé, 1959; Washburn, 1973; Harris et al., 1988; Rampton, 1988; French, 2007).

Some polygons, be they underlain at the margins by ice or sand, show neither elevated nor deflated margins. This is due to one of three conditions: **1)** wedge nascency, whereby marginal wedges have evolved insufficiently to show overburden uplift; **2)** truncated or stagnated growth, the result of thermal-contraction cycles having ended; or, **3)** a transitional stage between aggradation and degradation with the latter being insufficiently evolved for the margins to fall below the elevation of the centres.

*4.3 Thermokarst and ice-rich permafrost*

*Thermokarst* is a terrain type and a periglacial process (Harris et al., 1988). As the former, it references permafrost comprised of ice-rich sediments or *excess ice*. *Excess ice* references the volume of ice in the ground that exceeds the total pore-volume that the ground would have were it not frozen (Harris et al., 1988; also, see Taber, 1930; Penner, 1959; Rampton and Mackay, 1971; Washburn, 1973; Rampton, 1988; French, 2007).

Excess ice forms by way of ice *segregation*. Ice segregation, in turn, is the result of cryosuction pulling pore water to a freezing front where the ice consolidates interstitially into thin lenses and, over time, into more substantial and possibly tabular bodies of ice (Taber, 1930; Black, 1954; Penner, 1959; Rampton and Mackay, 1971; Rampton, 1988; French, 2007). Relatively fine-grained sediments, i.e. clays to silts to fine-sands, are particularly adept at hosting ice segregation (e.g. Washburn, 1973; French, 2017).

As the ice lenses aggrade, thermokarst terrain heaves; as the lenses degrade, the terrain settles (Taber, 1930; Penner, 1959; Hussey, 1966; Hughes, 1974; Rampton, 1988; Osterkamp et al., 2009; Farquharson et al., 2020). Hummocky and ice-rich permafrost often is indicative of ice depletion and may be due to mean-temperature disequilibrium within the region. However, the latter could also be connected with and the result of (larger-scaled) rises of mean temperature (e.g., Péwé, 1954; Czudek and Demek, 1970; Murton, 2001; Grosse et al., 2007; Osterkamp et al., 2009; Schirrmeister et al., 2013).

The time-frames of excess-ice aggradation and degradation, or of ice-induced heave and settlement, need not be proximal (e.g. Rampton and Mackay, 1971; Rampton, 1988; Farquharson et al., 2020). For example, most of the thermokarst lakes (filled with ice-derived meltwater) and alases (thermokarst-lake basins emptied of water by evaporation or drainage) in the Tuktoyaktuk Coastlands developed in the Holocene Era (e.g. Rampton and Mackay, 1971, Rampton 1988. Contrarily, the radiocarbon dating of wood ensconced in segregation-ice lenses and beds that are metres to tens of metres beneath the elevation datum of the region point to region-wide ice-enrichment having taken place thousands and possibly tens of thousands of years ago during the middle to late Wisconsinian glacial period (Rampton and Bouchard, 1988). This means that the geochronological offset of time between ice enrichment and depletion can be substantial, here, and possibly on Mars.

Thus, ice enrichment of the thermokarst-like terrain observed at the northern and southern mid-latitudes of Mars could have been enriched by the freeze-thaw cycling of water much earlier in the Amazonian Epoch than today, when boundary conditions were more clement; and, if the youthful mantle estimates at some locations on Mars are correct, then the ice-rich terrain could have been depleted by sublimation much later in the Amazonian Epoch, if not close to the present day, when thaw-associated boundary conditions at these locations seem improbable.

## 5. Possible periglacial landscapes in Protonilus Mensae

### 5.1 Clastically-sorted polygons?

In the case of the polygons and circular to sub-circular structures that populate the dark-toned terrain in unit *eHt* to the west of the geological contact separating it from unit *HNt, o*rigin cannot be deduced unambiguously from structure and form. However, the size, shape, networked distribution, bouldered margins and (presumed) sub-boulder sized centre-fills are distinctly similar to clastically-sorted circles observed in *wet* periglacial landscapes on Earth where the freeze-thaw cycling of water and cryoturbation take or have taken place **(Fig. 7a)** (also see Balme et al., 2009; Gallagher et al., 2011; Soare et al., 2014; Barrett et al. 2017). In as much as degraded basalts are widely present at the Martian mid-latitudes (e.g. Christensen et al., 2000; Poulet et al., 2007; Soare et al., 2015), it would not be implausible to ascribe a basaltic and relatively fine-grained composition to the centre fill of the candidate *CSCs* in unit *eHt*. This, too, would be in keeping, analogically, with the possibility of the candidate landforms on Mars being *CSCs*.

### 5.2 Impact craters

Other landforms in the dark-toned terrain show some morphological similarities with the candidate *CSCs* (see section 3.1). However, there are sufficient dissimilarities between these landforms and the candidate *CSCs* to discount a periglacial origin and sufficient similarities with small-sized impact craters to suggest synonymy with the latter **(Figs. 3b-e)**.

*5.3 Clastically non-sorted polygons and thermokarst-like depressions*

The polygons observed within the relatively light-toned surface material in the massif-centred basin exhibit size, shape, networked distribution and margins that are consistent with polygons formed by thermal-contraction cracking in permafrost regions on Earth (e.g. Lachenbruch, 1962; Washburn 1973; French, 2017) **(Fig. 7b)** and, it is thought, elsewhere on Mars (Pechmann, 1980; Costard and Kargel, 1995; Seibert and Kargel, 2001; Morgenstern et al., 2007; Soare et al., 2008; Levy et al., 2009a,b; Séjourné et al., 2011, 2012; Oehler et al., 2016).

The origin of thermal-contraction polygons is rooted in cyclical and sharp drops of below zero temperatures in permafrost (Lachenbruch, 1962). The related stresses, strains and relaxation associated with these cycles occur regardless of whether the affected terrain is ice-rich or ice-poor (Lachenbruch, 1962). Water undergoing cyclical changes of phase is not a requirement of this process, or of the derivative formation of polygonised terrain.

As seen above, polygonised, clustered and irregularly-shaped depressions that are rimless and metres- to decametres-deep also punctuate the relatively light-toned surface material, here **(Fig. 6)** and throughout the mid-latitudes of the northern plains (e.g. Costard and Kargel, 1995; Morgenstern et al., 2007; Soare et al., 2007, 2008; Lefort et al., 2009; Ulrich et al., 2010; Séjourné et al., 2011, 2012; Dundas et al., 2015; Barrett et al., 2017; Dundas, 2017). Often described as thermokarst, these structures are deemed to be akin to thermokarst on Earth and are assumed to comprise excess ice. When *NSPs* are observed in their midst, regardless of whether the polygons show high or low centres, there is a relatively high degree of probability, based once again on candidate Earth analogues, that the surface and near-surface material are ice rich (e.g. Costard and Kargel, 1995; Morgenstern et al., 2007; Soare et al., 2007, 2008; Lefort et al., 2009; Ulrich et al., 2010; Séjourné et al., 2011, 2012; Dundas et al., 2015; Barrett et al., 2017; Dundas, 2017).

*5.3.1. Excess ice origin?*

Currently, liquid water is not stable with regard to the triple point at the mid- to -higher-latitudes of Mars (Mellon and Jakosky, 1993, 1995). This would seem to discount the plausibility if not the possibility of the candidate thermokarst-landforms having developed by the freeze-thaw cycling of water (Morgenstern et al., 2007; Lefort et al., 2009; Cull et al., 2010; Ulrich et al., 2010; Séjourné et al., 2010, 2011; Dundas et al., 2015; Dundas, 2017). In its place, ice-enrichment hypotheses tend to invoke *dry* processes such as those involving adsorption-diffusion cycles (e.g. Mellon and Jakosky, 1993, 1995; Mellon et al., 2004. Morgenstern et al., 2007; Lefort et al., 2009; Dundas et al., 2015; Dundas, 2017).

We acknowledge and agree that the simplest and most plausible way to explain the devolatilization of thermokarst under current or recent conditions would have to be by sublimation. Similarly, adsorption-diffusion cycles would seem to be the most plausible means to explain the ice enrichment or volatilization of permafrost. On the other hand, the geothermal gradient in the sub-surface precludes ice-adsorption below a skin depth of a metre or so; in addition, adsorption saturates near-surface pore space early on in the process; this forms an impermeable barrier below which no further adsorbed ice can develop (e.g. Clifford, 1993; Mellon and Jakosky, 1993, 1995). By default, the iterative or episodic freeze-thaw cycling of water is the only widely-recognised process by which ice-enrichment to decameters of depth, a depth that is not unusual for thermokarst-like depressions in regions such as Utopia Planitia (e.g. Morgenstern et al., 2007; Séjourné et al 2011), can take place.

Three further points also favour the freeze-thaw cycling hypothesis. First, ice-enrichment and ice-depletion need not be coeval. Some thermokarst landscapes on Earth (see section 4.3) show offsets of tens of thousands of years between the periods of ice aggradation and degradation. Similarly, ice-enrichment on Mars could have preceded its depletion by far, occurring at a time when water was more stable at or near the surface than today. Second, as long as water is available and boundary conditions are appropriate the development of thermokarst by freeze-thaw cycling could occur quickly, as it does on Earth, occasionally.

For example, within five years of a thermokarst lake having been drained artificially in the Tuktoyaktuk Coastlands on Earth, ice wedging, polygonization and nascent pingo formation were observed (e.g. Mackay, 1997; Mackay et al., 2002). The geographical reach of meta-stable regions at the middle latitudes could well have wandered stochastically throughout the Mid Amazonian Epoch (e.g. Haberle et al., 2001; Hecht, 2002), by way of their geographical reach and their temporal span. Third, this would be the case especially were near-surface perchlorate brines present (e.g. Gallagher et al., 2011; Barrett et al., 2017; 2018; Soare et al., 2018, 2021a; also, see brine references in, e.g. Renno et al., 2009; Martinez et al., 2017; Primm et al., 2019; Chevrier et al., 2020).

## 6   Glacial landscapes on Earth

### *6.1 Glacial ice, cirques, flows, debris aprons and moraines* **(Fig. 8)**

*Glacial* ice, be it within an ice sheet, ice cap or mountain glacier, accumulates by iterative ice/snowfall deposition, the subsequent burial, compaction and recrystallisation of which generates its primary structure (Jennings and Hambrey 2021). Secondary structure, manifesting as folds, foliation, and crevassing, is produced as deep ice undergoes ductile deformation (Hambrey and Müller, 1978) or shallow ice undergoes brittle deformation and fracture (Colgan et al., 2016).

Gravity and the internal deformation of the ice are the principal mechanisms of glacial ice-*flow,* irrespective of size and thermodynamic state (e.g. Cuffey and Paterson, 2010; Barry and Gann, 2011). *Cirques* are amphitheatre-like erosional hollows or scars, presently or formerly occupied by glacial ice at or close to mountain summits, and are characterized by steep headwalls and over-deepened floors (Barr and Spagnolo 2015) **(Figs. 8a-b)**. Geographically, they can mark the origin of glacial flow.

At lower (relative) elevations glacial-flow surfaces and *margins* can demarcated by debris-laden ridges or moraines (e.g. Martini et al., 2001) **(Fig. 8a)**. *Moraines* are created and modified by a range of processes that include but are not limited to: bulldozing/pushing and gravity-driven movements (e.g. Benn and Evans, 2010). Generally speaking, moraine types are characterized by their location within or adjacent to flow surfaces and bodies at the fore, side or in the midst of these surfaces.

For example, terminal moraines delimit the maximum horizontal extent of a glacier. They are composed of till and reworked stratified material, form at the front of actively moving glaciers or of stagnant ice, and are curvilinear or lobate (e.g. Martini et al., 2001). Recessional moraines form on the lee or background of the terminal moraines. They are younger, sometimes serialized and often less massive than terminal moraines. They form to the lee or in the background of terminal moraines as their recession pauses or stands-still (e.g. Hambrey, 1994). Other moraine types include lateral moraines, framing glacial flow on either of its sides normal to the flow front; medial moraines, occurring where lateral moraines merge at the confluence between ice-flow units; and, ground moraines, i.e. low-relief and topographically uneven terrain deposited by retreating glaciers (e.g. Hambrey, 1994; Martini et al., 2001).

Where debris sources from adjacent topography, i.e. valley walls, is particularly high, debris-covered glaciers may develop. Debris cover above a threshold minimum-thickness acts to retard melt-rates, dampening the response of these features to warming climate (e.g. Anderson and Anderson 2016, Immerzeel et al., 2020).

Morphologically, glacial cycles end with a mass loss by ablation and the fragmentation of ice deposits.

*6.2 Glacial landscapes in Protonilus Mensae?*

Numerous surface features radial to the massifs within the central basin of our study region, discussed above (see section 3.2), conform morphologically, geographically, and in their spatial association with glacial landscapes on Earth and no less so with candidate glacial-landscapes elsewhere on Mars (e.g. Souness and Hubbard, 2013; Hubbard et al., 2014; Baker and Head, 2015; Brough et al., 2016; Hepburn et al., 2020a, b; Gallagher et al., 2021).

Collectively, the term *viscous-flow features* [*VFFs*] refers to the group of constrained surface materials whose (topographically) draping planform, slope angles and consistency with the flow laws of ice point to ice-based viscous deformation possibly on Mars as on Earth (e.g. Milliken et al., 2003). Globally, *VFFs* are characterized by muted (underlying) terrain and adjacency to massifs, scarps or crater walls (e.g. Levy et al., 2009a; Milliken et al., 2003; Hepburn et al., 2020a). Some *VFFs* are incised by longitudinal and/or transversal fractures (e.g. Mangold et al., 2003; Pedersen and Head, 2010; Hubbard et al., 2014) and/or polygonised terrain (e.g. Levy et al., 2009a; Sinha and Murty, 2015; Soare et al., 2021c). Where ice-loss or ablation is thought to have occurred, *VFFs* are discontinuous, morphologically irregular (e.g. Milliken et al., 2003; Levy et al., 2009a; Pedersen and Head, 2010; Brough et al., 2016) and show decametres-scale patches of small ridge/trough assemblages or *brain terrain* (e.g. Levy et al., 2009a).

Originally, only features observed debouching from alcoves were termed *VFFs* (Milliken et al., 2003). However, the definition has since been revised and, following Souness et al. (2012), we use *VFF* as an umbrella term encompassing a range of landforms subdivided according to their size and context. Two types of *VFFs* are particularly relevant to our work:

a) *Glacier-like forms* [*GLFs*] are the lowest order form of *VFFs* and are similar in planform appearance to valley glaciers or debris-covered glaciers on Earth (Souness et al., 2012). They originate in *cirque*-like alcoves at or near glacier summits, funnel through narrow valleys and are demarcated downslope by *moraine-like ridges* [*MLRs*] (e.g. Arfstrom and Hartmann, 2005; Pedersen and Head 2010; Souness and Hubbard, 2013; Sinha and Murty, 2015; Brough et al., 2016).

b) *Lobate debris-aprons* (*LDAs*) are larger *VFFs* which demarcate the collective distribution of flow, be it continuous or discontinuous, from the summit or near-summit cirques through to marginal, terminal or recessional moraine-like ridges (e.g. Souness and Hubbard, 2013; Brough et al., 2016; Hepburn et al., 2020b). Underlying or buried ice may be present, stabilised by debris or a sublimation lag (e.g. Mellon and Jakosky, 1993, 1995; Milliken et al., 2003; Levy et al., 2009a; Pedersen and Head, 2010; Hubbard et al., 2014; Baker and Head, 2015; Sinha and Murty, 2015; Hepburn et al., 2020b).

## 7. The periodicity of *icy* and of *ice-rich* landscapes?

As noted above (see Introduction), there is general agreement that the mid- to high latitudes of the northern hemisphere of Mars are draped by an atmospherically-precipitated and metres-thick mantle(s); the mantle(s) is/are thought to be composed of icy and/or ice-rich material and/or a sublimation lag (e.g. Mellon and Jakosky, 1995; Mustard et al., 2001; Milliken et al., 2003; Head et al., 2003; Forget et al., 2006; Dickson et al., 2008; Madeleine et al., 2009; Souness and Hubbard, 2013; Baker and Head, 2015; Soare et al., 2021a, b).

On Earth, glacial ice and the landscapes derived therefrom, require no phase transition of water to develop. By contrast, ice-rich (permafrost) landscapes comprise excess ice, which is interstitial. Interstial ice, as discussed above, requires meltwater migration into the pore space of the host material to develop.

On Earth, particularly during the Quaternary Period, glacial/deglacial (or periglacial) cycles occurred regularly, as have associated variances in regional and/or global mean-temperatures. Below, we use the Tanaka et al. (2014) description of the geological units that frame our study region and crater-size frequency distributions to contextualise the possibility that the intertwining of periglacial and glacial periods is no less present on Mars than on Earth, at least during the Mid to Late Amazonian Epochs. We also suggest that this intertwined periodicity extends far more deeply into the history of Mars than has been shown hitherto in the literature.

## 8. Age-dating of the *icy* vs *ice-rich* landscapes in our study region

*8.1 Age estimates of morphologically similar terrain elsewhere*

Crater-based age estimates of the *VFFs* at or near the Mars dichotomy describe a temporal reach from the recent past back through to the mid-Amazonian Epoch i.e. ~1 - ~100 Ma (e.g. Head et al., 2003; Morgan et al., 2009; Souness and Hubbard, 2013; Hubbard et al., 2014; Sinha and Murty, 2015; Brough et al., 2016); ~100 Ma - ~1 Ga (e.g. Morgan et al., 2009; Baker et al. 2010; Sinha and Murty, 2015; Butcher et al., 2017, 2021) and, perhaps, even earlier than that, ~1 Ga (Levrard et al., 2004).

By means of contrast, most ~~crater-based~~ age estimates of possible periglacial landscapes inclusive of *NSPs* and thermokarst-like depressions at/near the Mars dichotomy or at the mid- to high- northern latitudes, show relatively short and youthful age ranges: <~0.1 Ma (Mustard et al., 2001; also, Milliken et al., 2003); ~0.1 Ma to ~1 Ma (Levy et al., 2009b; Mangold, 2005); ~0.4 - ~2.1 Ma (Head et al., 2003); ≤~3.0 Ma (Kostama et al., 2006). Recently, Soare et al. (2020) reported a minimum age-estimate of ~100 Ma for possible periglacial terrain at the mid-latitudes of Utopia Planitia and immediately to the north of the Moreux impact-crater. Exceptionally, small-sized outcrops of possible thermal-contraction polygons thought to have formed in the Hesperian Epoch have been observed at the Gale Crater (Le Deit et al., 2013; Oehler et al., 2016).

Most of the surface-age estimates associated with candidate *CSCs* reported elsewhere, as with the *NSPs*, are youthful: ~0.1 Ma, at the high latitudes of the Heimdal Crater (Gallagher et al., 2011); and, ~2.0 - ~8.0 Ma, at the near-equatorial latitudes of Elysium Planitia and Athabasca Valles (Balme et al., 2009, inferred from Burr et al., 2005). Below, we show that the candidate *CSCs* in unit *eHt* could be 1 - 2 orders of magnitude older than the candidate *CSCs* referenced above.

*8.2 Age estimates of units eHt and HNt*

Absolute model ages of units *eHt* and *HNt* are estimated to be 3.59 – 3.69 Ga and 3.70 – 3.99 Ga respectively by Tanaka et al (2014) assuming the chronology model of Ivanov (2001) **(Fig. 9)**. These ages represent extensive units mapped at a global scale with ages based on several crater-count locations in widely disconnected and disparate areas. However, they are generally consistent with our observations of the local underlying material.

To the northeast of *HiRISE* image ESP_028457_2255 four craters are located within a highly-localised topographical depression **(Fig. 10a, also see Fig. 9 and 11a)**, so named because of its abrupt loss of elevation, i.e. ~70 m **(Fig. 11a)**. The elevation loss occurs where unit *eHt* is thin and/or discontinuous **(Fig. 10a; also Fig. 2)**. It is reasonable to assume that these craters incise the underlying basement unit, presumably *HNt* in the Tanaka et al. (2014) map; this is consistent with a model age of ~3.7 Ga.

The largest crater in *CTX* image F21_044083_2248_XI_44N317W has a diameter of ~5 km **(Fig. 2)**. A single crater of this size within the area represented by the *CTX* image is consistent with the model crater-retention age of a divot-like topographic depression **(Fig. 10)**, as a crater of this size would be predicted to form within ~3.5 Gyrs using the Hartmann 2005 model, or ~3.7 Gyrs using the Ivanov (2001) model.

*8.3 The dark-toned terrain: impact cratering and age estimates*

The population of candidate impact-craters in the dark-toned terrain, immediately adjacent and to the west/north-west of the geological contact separating units *eHt* from *HNt,* was catalogued **(Fig. 12)** and the crater size-frequency distributions (*CSFDs*) compared with model crater-retention age isochrons **(Fig. 13)**. The depressions with diameters < 80 m were broadly classified based on morphology as *Types 0 - 3*, from least-likely to be impact related to most-likely impact related. Larger unambiguous craters were identified as well as apparent buried and ghost craters.

*Type 0* depressions are shallow, often irregular or elliptical in planform, with no sharply defined edge **(Fig. 3b)**. An impact origin for these depressions is highly unlikely. These features were excluded from evaluation and are not included in our figures or discussion. *Type 1* depressions are circular in planform giving them the appearance of a heavily-eroded crater and they could be impact-related **(Fig. 3c)**. However, their muted topographic expression and lack of a sharply-defined edge makes their identification as remnant impact craters ambiguous given the overall texture of the surrounding terrain. *Type 2* depressions are circular with uplifted rims and steeper interior wall slopes than typical of the surrounding depressions, i.e. possible *CSCs,* making them strong candidates for degraded, remnant impact craters (**Fig. 3d**). *Type 3* depressions are bowl-shaped with sharp edges or rims and steep inward slopes (**Fig. 3e**). They are all smaller than ~50 m and many occur in tight clusters. These are confidently Identified as relatively-fresh impact craters that have experienced minimal post-formation modification.

Where they are distributed densely they resemble crater clusters formed by the fragmentation of impactors during passage through the Martian atmosphere (e.g. Ceplecha et al., 1998; Artemieva and Shuvalov, 2001; Popova et al., 2003; 2007; Williams et al., 2014) as commonly observed among the population of newly formed craters identified by *CTX* image temporal pairs throughout the *MRO* mission (Daubar et al., 2013, 2019).

At larger scales, >~80 m, impact craters are confidently identified due to their size, even when heavily modified or filled (**Fig. 4**), as their size exceed the characteristic length-scale of the terrains polygonal texture. These craters can be either *filled* or *unfilled*. The unfilled craters incise the current surface and likely were formed after the lithic unit was emplaced as they retain a bowl-shape with little infilling material. A subset of these craters appear to have topographic benches outlining the lower portion of the crater interiors (**Figs. 4a-b**). This could result from a transition in target properties and represent a stratigraphic horizon at depth (e.g. Oberbeck and Quaide, 1967; Prieur et al., 2018; Martellato et al., 2020). These craters have a narrow range of diameters, 106 - 126 m, suggesting the transition occurs at a depth ~20 m.

The largest crater in the population ($D$ = 350 m) has a clearly identifiable edge and ejecta material that appears to superpose the surrounding terrain (**Fig. 4b**). This crater likely post-dates the emplacement of the terrain rather than extending through from beneath or being embayed. Though the crater interior has accumulated material, it does not contain the same polygonal morphology as the lithic unit and the rim remains well preserved and exposed. The abrupt truncation of polygons of the dark-toned unit at the ejecta edges also suggest the ejecta overlays the polygons. The overall topographic relief of the crater and its ejecta is in stark contrast to an observed population of subdued, shallow circular depressions typically ∼>100 m with arcuate ridges, fractures, and scarps (**Fig. 4d**). These are interpreted to be ghost craters representing the pre-existing craters on the older underlying surface.

A smaller class of circular features, frequently tens of meters in diameter, has also been identified with central mounds forming a circular moat (**Fig. 4c**). If these represent impact craters, these would have formed prior to the emplacement of the current surface materials and could thus be synformational craters embedded within the volume of material.

The differential *CSFDs* of these different classes of craters are plotted in (**Fig. 13**). The *Type 3* craters (**Fig. 13d**) plot near the 1 Ma isochron for $D > 10$ m suggesting the current surface has not experienced substantial modification in the last ∼1 Ma. The *roll off* at smaller diameters usually is observed in *CSFDs* as the crater diameters approach the image resolution limit. The *CSFD* of the *Type 2* craters (**Fig. 13c**), which have a more degraded appearance relative to the *Type 3* craters, is between the ∼10 Ma and ∼100 Ma isochron at $D > 20$ m and shallows in slope at smaller diameters suggesting surface modification has preferentially removed the smaller craters from the population (e.g. Öpik 1965; Chapman et al., 1969; Hartmann et al. 1971; Smith et al., 2008; Williams et al., 2018; Palucis et al., 2020). The *Type 1* craters (**Fig. 13b**) have a peak in crater density at $D \sim 10 - 40$ m with a steeper *CSFD* slope at $D > 20$ m than the model isochrons. At smaller diameters ($D \lesssim 10$ m) there is a downturn in the *CSFD* down to $D \sim 6$ m before increasing again at smaller diameters. Since the texture of the terrain occurs at this length-scale, and their morphology made the identification of the origin of the *Type 1* craters ambiguous, this suggests that many of the features in this category have been misidentified. Thus this class of features has been excluded from further consideration. However, their exclusion makes little difference on the age interpretation.

There are 13 large, $D > 80$ m, unfilled craters which, due to their size and depth, are confidently identified as impact craters. These craters, along with the *Type 2* and *3* craters, provide a total population of 404 features confidently identified as impact craters. The combined differential and cumulative *CSFDs* are shown in (**Fig. 14)**. The largest craters suggest the age of the dark-toned terrain is >~100 Ma with the largest crater, $D = 350$ m, expected to form on a surface >~1 Ga. The age implied by the single large crater should be viewed with caution as dating surfaces using just a single, or a few large craters, can lead to erroneously old model ages and uncertainties in model surface ages grow with smaller areas due to a loss in statistical precision (e.g. van der Bogert et al., 2015; Warner et al., 2015; Palucis et al 2020). However, given the overall population of craters, it is unlikely the age of dark-toned terrain is younger than <~100 Ma although it could be as old as ~1 Ga.

*8.4 The light-toned terrain: impact cratering and relative age estimates*

At the geological contact to the west of the basin-centred massifs in *HiRISE* image ESP_028457_2255 some of the light-toned moraine-like ridges intercept unit *eHt* and, seemingly, have piled up at disparate contact locations **(Fig. 5d)**. This suggests that the moraine-like structures post-date unit *eHt*. Based on age estimates of the light-toned terrain this means that the *MRLs* and other candidate glacial features are ~10 - ~100 Ma **(Fig. 15)**. This is consistent with some of the other estimates of possible glacial landscapes in the region (e.g. Head et al., 2003; Morgan et al., 2009; Souness and Hubbard, 2013; Hubbard et al., 2014; Sinha and Murty, 2015; Brough et al., 2016).

However, using the sharply-delineated contact between unit *HNt* (possibly comprised of degraded or relict glacial material) and unit *eHt* (possibly composed of ice-rich material) as a putative terminus **(Figs. 11a-b)** and an assumed (basin-floor) flat-bed, we applied a 2D perfect plasticity model of glacial flow to the observed *LDA* in the massifs-centred basin **(Fig. 11b)**. We found that the modelled profile is a poor fit for the measured profile of the *LDA* and a thicker ice mass is predicted based upon our initial assumptions. The discrepancy between the measured and modelled profile suggests one of three things:

1)  Our assumed *LDA* terminus is incorrect and the true (buried and underlying) *LDA* terminus extends beyond the apparent visible-boundary contact beneath the dark-toned terrain;

2)  Our flat-bed assumption is a poor representation for the underlying topography; or,

3)  The *LDA* (and *GLF*) surfaces have deflated significantly since a previous glacial maximum.

We cannot rule out **2)** without *SHARAD* but based on the surrounding terrain this seems unlikely. The massif may not transition to a vertical profile at the intersection with the *LDA*, and the gently sloping profile of the massif here may hint at its continuation beneath the *LDA*. However, extrapolating the unknown massif topography from the visible topography is unlikely to affect our modelled profile because the model initialises at the putative terminus and propagates up-glacier. Changing bed topography towards the upper margin of the *LDA* would have no effect on the shape of the profile prior and the profile overall would remain concave. We rule out **3)** as this would be inconsistent with the observed interception of the darker-toned terrain by the lighter-toned (possible) terminal or push moraines. Moreover, models of perfect plasticity used elsewhere in the contiguous Deuteronilus-Protonilus Mensae regions are a good fit for contemporary lobate debris-apron surfaces (e.g. Karlsson et al., 2015; Schmidt et al., 2019).

If the unobserved *LDA* extends beyond the geological contact separating unit *eHt* from unit *HNt* and underlies the former, then we can infer that it predates unit *eHt* and must be ≥ ~100 Ma and, possibly, ~1 Ga. This would also suggest that the *LDA* and *GLF* frame or bracket unit *eHt*, stratigraphically and temporally.

The massifs-centred basin, as discussed above, also hosts (clastically non-sorted) polygonised terrain punctuated by thermokarst-like depressions. Wherever these assemblages are observed the texture of the terrain incised by them is relatively smooth (at least at the *HIRISE*-scale of resolution) and the underlying topography is muted.

This could be the result of being nested within atmospherically precipitated and relatively recent icy or periglacially-revised ice-rich terrain, i.e. ~10 - ~100 Ma **(Fig. 15)**. Similar albeit slightly more youthful age estimates of mantled terrain have been reported elsewhere at the mid- to higher-latitudes of the northern plains (e.g. Mustard et al., 2001; Head et al., 2003; Milliken et al., 2003; Mangold et al., 2005; Levy et al., 2009b; Mangold, 2005; Kostama et al., 2006). This would also suggest that the observed and buried/unobserved but hypothesised *VFFs* in the massifs-centred basin of unit *HNt* constitute a temporal or geochronical gap that separates the formation age of the candidate *CSCs* in the relatively dark-toned terrain of unit *eHt* and the *NSPs* in the former.

## 9. Discussion & Conclusion

### 9.1 Relative stratigraphy and geochronology

Based on the putative observation of *VFFs* at the surface of the massifs-centred basin in unit *HNt* and the hypothesized presence of buried *VFFs* on the floor of this basin, we surmise the presence of (at least) two stacked and temporally-distinct periods of glacial activity within the basin.

We suggest that the reach of the modelled but buried and unobserved *VFFs* extended to the geological contact separating unit *HNt* from unit *eHt* and beyond, geographically. In this regard, we propose that the inward-oriented benches or terraces observed within some of the candidate impact craters in unit *eHt* exhume unit *HNt*. If so, then this would suggest two things. First, unit *eHt* and the putative *CSCs* that incise it overlie unit *HNt* and, derivatively, postdate it, as Tanaka et al. (2014) have argued. Second, our crater-based age estimates of the observed, surface *VFFs* of unit *HNt* point to a formation age that is younger, substantially so, than unit *eHt*.

### 9.2 Principal findings

As such,

1) If the periglacial categorisation of the *CSCs* is correct and were the min/max age estimates (~100 Ma - ~1 Ga) of the dark-toned terrain incised by them valid, then the *CSCs* would comprise the oldest *sorted* periglacial features reported in the literature.

2) If the periglacial categorisation of the *NSPs*/thermokarst-like depressions is correct, and were the min/max age estimates (~10 Ma - ~100 Ma) of the light-toned terrain incised by them valid, then the range of age that separates the periglacial landscapes comprised of the *CSCs* and the NSPs/thermokarst like features would be much greater than at any other location reported in the literature.

3) The temporal intertwining of the two proposed glacial periods (based on the observed and modelled flow of the massif-centred *VFFs* in our study region) amidst the two proposed periglacial periods (based on the age estimates of the dark and light-toned terrains) comprises a ~1 Gyr reach into the Amazonian Epoch and its paleo-climatic record.

4) The last point highlights the extent to which relative stratigraphy, tied to crater-based age estimates, can be used to identify the cyclicity if not the alternance of glacial/deglacial boundary conditions in Protonilus Mensae and, perhaps, elsewhere on Mars through a significantly long period of the planet's late geological history.

**Acknowledgements and funding**

We are grateful for the extremely positive and supportive comments made by each of the two anonymous reviewers of this chapter. FEGB is part of the PALGLAC team of researchers and received funding from the European Research Council (ERC) under the European Union's Horizon 2020 research and innovation programme (Grant agreement No. 787263).

**Figures**

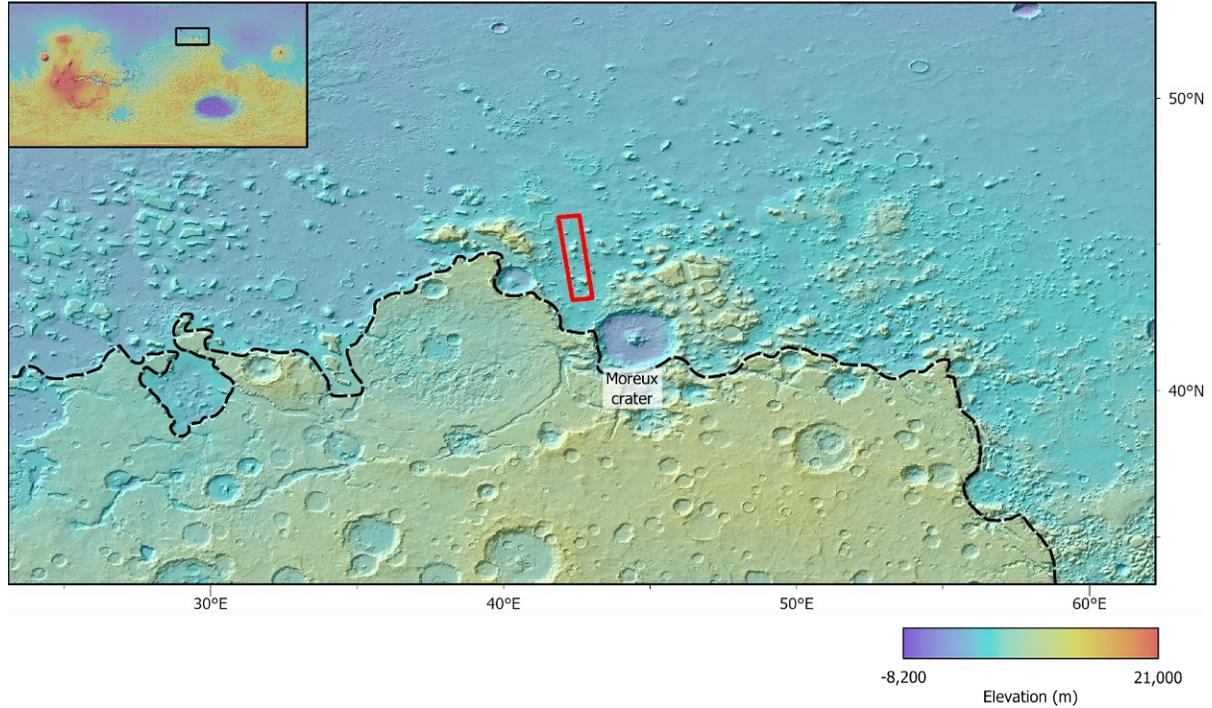

**Fig. 1:** The geographical footprint of our study area (red rectangle) in the *P*rotonilus *M*ensae [*PM*] region of Mars (extent shown by black box in inset of global elevation map of Mars). The black dashed-line highlights the Mars crustal dichotomy and the proximity of our footprint to it. Background colour comprises *MOLA* global-elevation (Zuber et al., 1992) in an equirectangular projection. *MOLA* data credit: *MOLA* Science Team, Arizona State University.

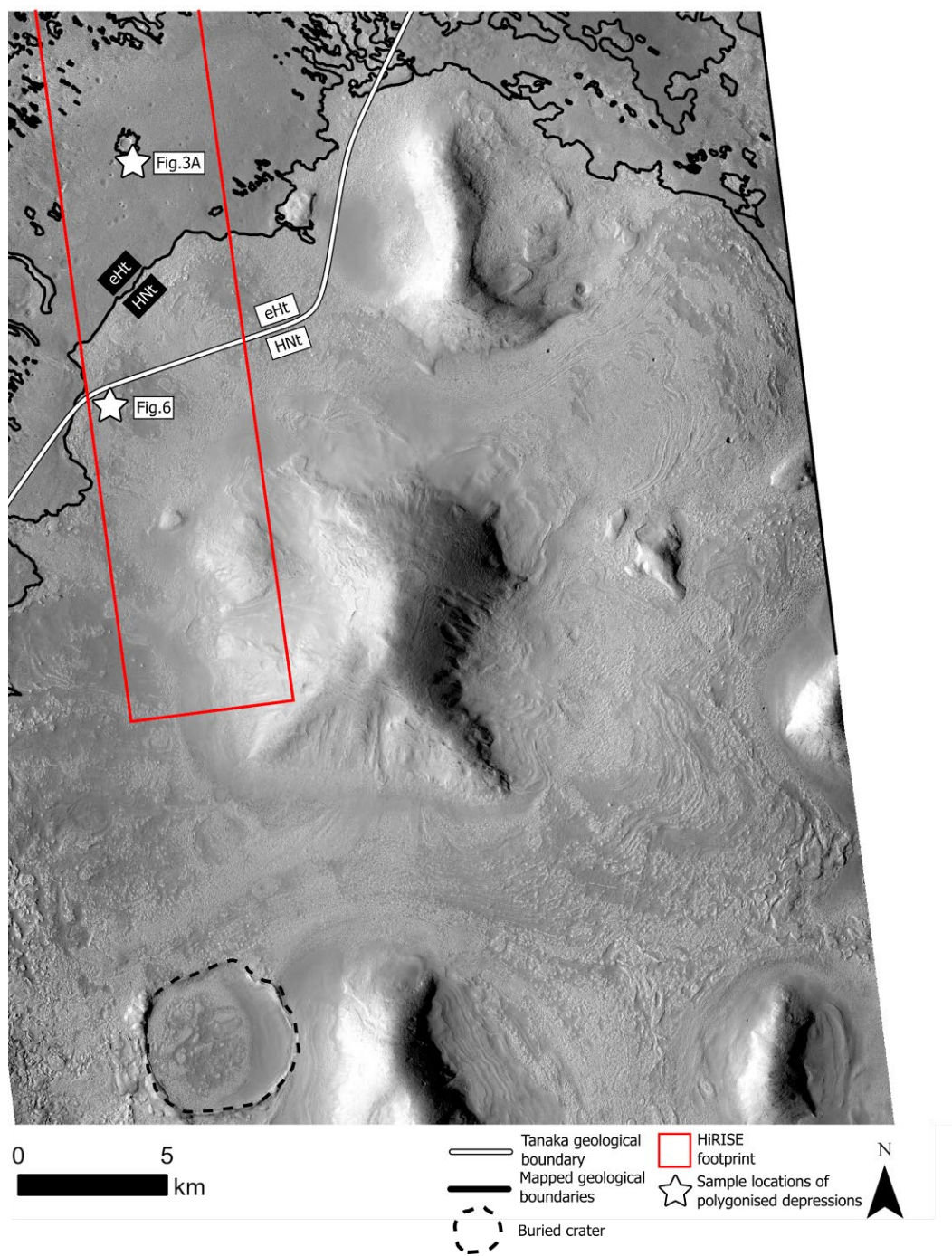

**Fig. 2:** C*TX* image F21_044083_2248_ XI_44N317W of geological units *eHt* and *HNt* in the *PM* region. The two units are separated by a contact first identified by Tanaka et al. (2014) and refined, here. The white line coincides with Tanaka's original boundary, derived of a large regional-scale map; the black line marks the updated contact. Age estimates of the large

crater (serrated circle) suggest that it intercepts the floor of *HNt* at depth (Soare et al., 2022b). The red rectangle represents the footprint of *HiRISE* image ESP_028457_2255. Stars mark the sample locations of (candidate) clastically-sorted circles in unit *eHt* (see **Fig. 3a)** and polygonised but not clastically-sorted thermokarst-like depressions in unit *HNt* (see **Fig. 6)**. North is up. *CTX* image credit: *NASA/JPL*/Arizona State University. *HiRISE* image credit: *NASA/JPL*/University of Arizona.

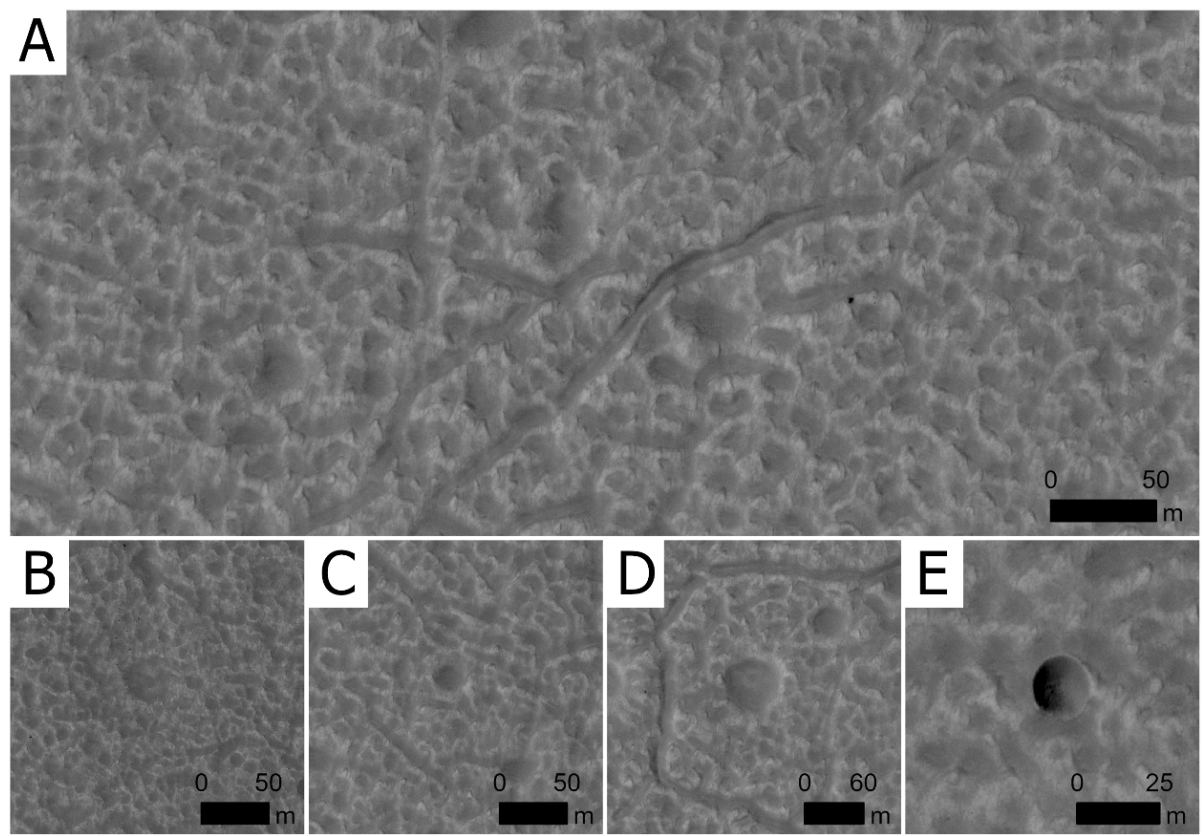

**Fig. 3: a)** Example of ubiquitous surface-coverage of unit *eHt* by decametre-scale circular to sub-circular or quasi-polygonised structures, elevated at the margins. The margins are punctuated by boulders and show a slightly lighter tone than the terrain circumscribed by them. *HiRISE* image ESP_028457_2255. Examples of four morphologic categories of depressions based on similarities to impact craters. **b)** *Type 0 - unlikely*: shallow, often irregular, or elliptical in shape with no apparent rim. **c)** *Type 1 - possible but ambiguous*: similar to *Type 0* but are circular in planform making them candidates for being impact related. **d)** *Type 2 - probable*: circular with uplifted rims and steeper interior wall slopes than typical of the surrounding depressions. **e)** *Type 3 - unambiguous*: bowl-shaped with sharp edges or rims and steep inward slopes. Scale bars are 20 m. North is up in all panels.

*HiRISE* ESP_028457_2255. North is up in all panels. Image credit: *NASA*/*JPL*/University of Arizona.

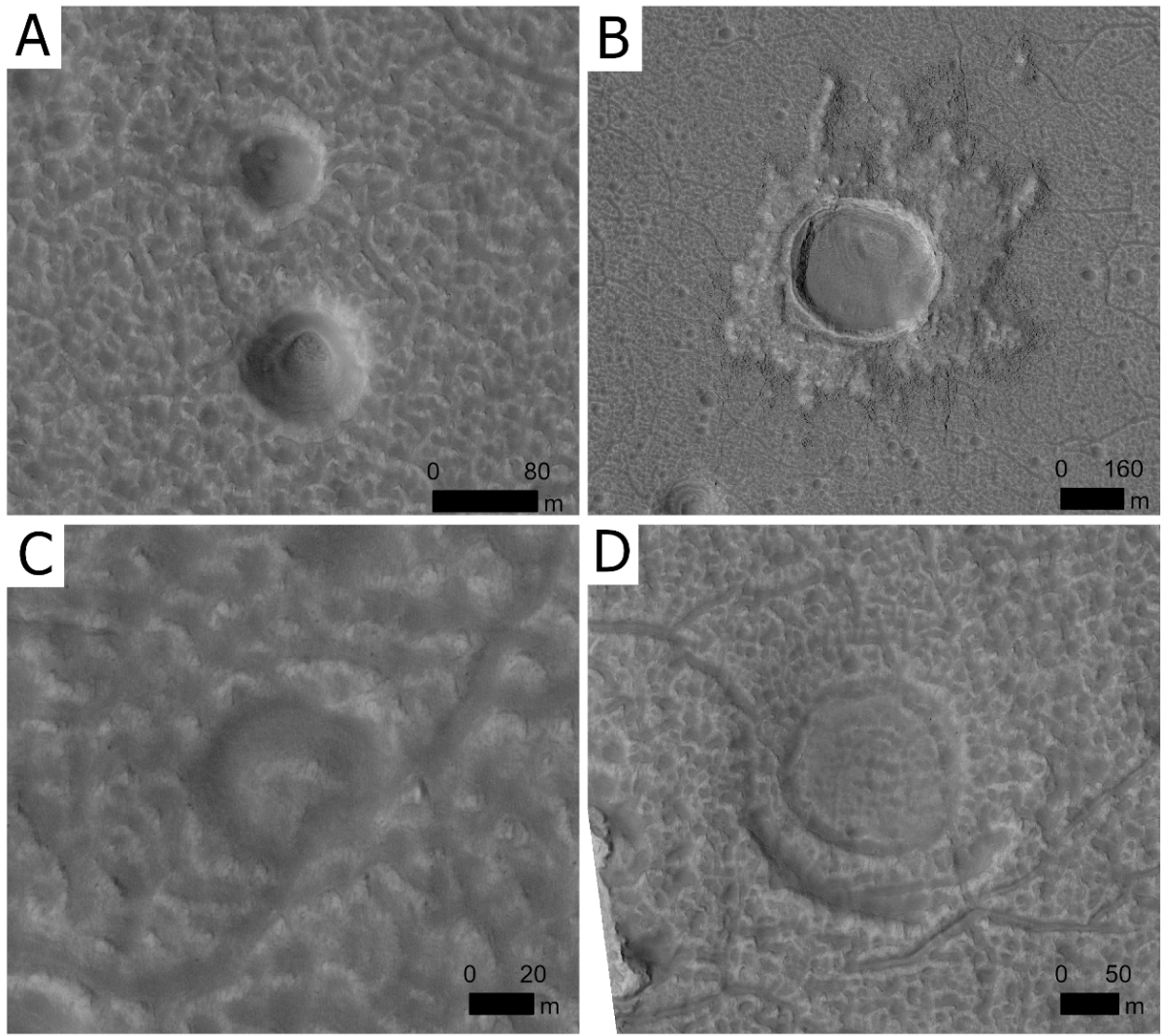

**Fig. 4: a)** Larger, unfilled bowl-shaped depressions confidently identified to be impact craters due to their size and depth. The lower crater is an example of a subset of these craters that have terraces in the interior wall outlining their center near the floor, possibly resulting from a transition in target properties. **b)** Largest crater ($D = 350$ m) identified in the count region with a clearly visible rim and surrounding ejecta material that appears to overlay the adjacent terrain. **c)** Example of a class of depressions with central mounds that may represent buried impact craters formed prior to the emplacement of the current exposed

surface materials and could thus represent embedded craters exposed by exhumation. **d)** Class of subdued, shallow circular depressions with arcuate ridges, fractures, and scarps typically $D > 100$ m. These may represent ghost craters from a preexisting population of craters on an older underlying surface. North is up in all panels. *HiRISE* image ESP_028457_2255. Image credit: *NASA*/*JPL*/University of Arizona.

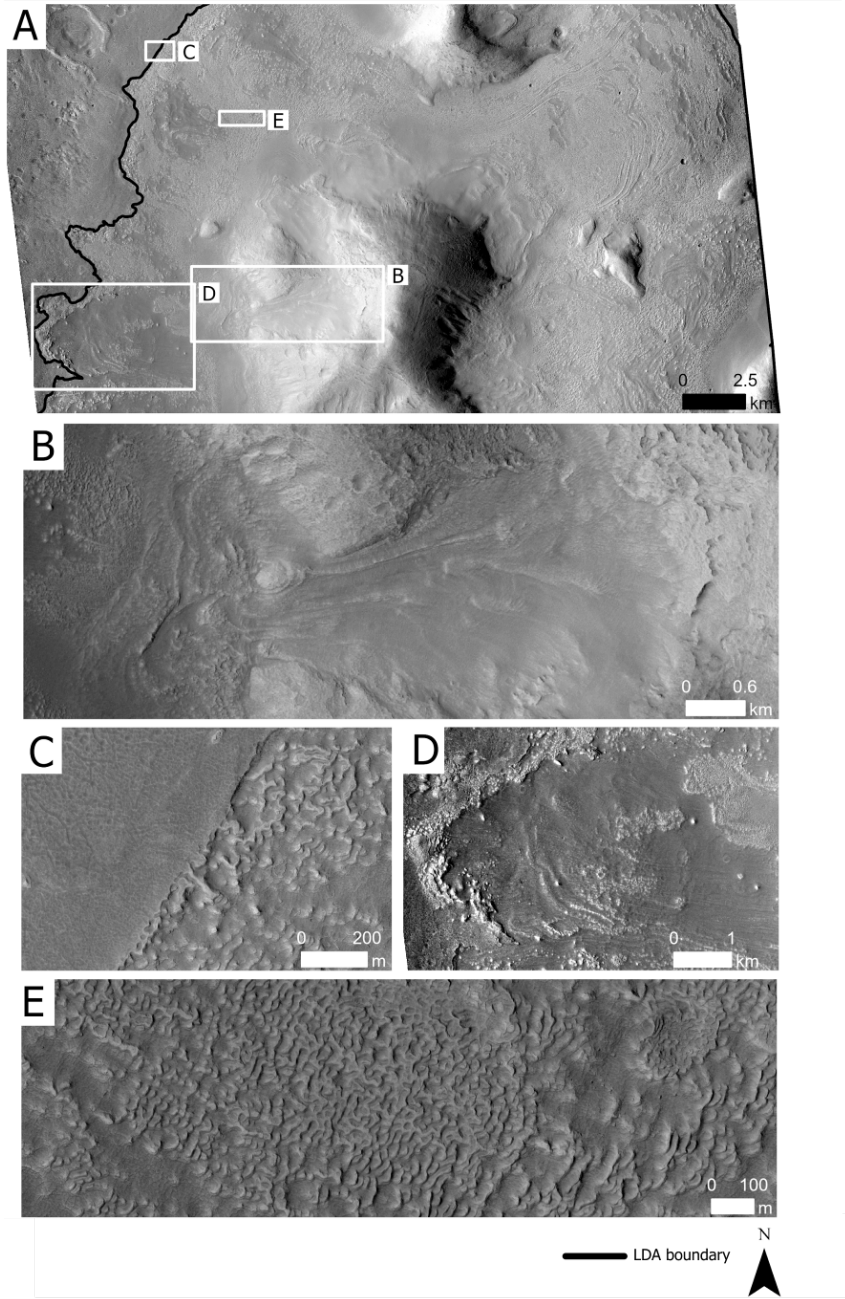

**Fig. 5: a)** Magnification of *CTX* (massifs-centred) context image. The black line demarcates the western margin of the principal *l*obate-*d*ebris *a*pron [*LDA*] in the image. **b)** Amphitheatre-shaped depression heading possible *VFFs* through a valley and towards a series of moraine-like ridges [*MLRs*] on the valley floor. Note possible medial *MLRs* in the midst of the *VFFs*. **c)** Degradational contact between the dark-toned terrain and the *LDA*. The degradational

contact appears to have eroded backwards, revealing underlying textures consistent with the upper reaches of the *LDA*. **d)** A series of candidate push moraines associated with the alcove sourced glacier-like form. The possible moraines appear to pile up at the contact with unit *eHt* and the topographical profile of this location indicates that the former are at a higher elevation than the latter (see **Fig. 11a)**. This would be consistent with *CSFD*-based age estimates suggesting that unit *eHt* predates the light-toned surface of unit NHt. **e)** Small-sized ridge/trough assemblages that are open or closed, possibly formed by ablation and/or devolatilization and erosion.

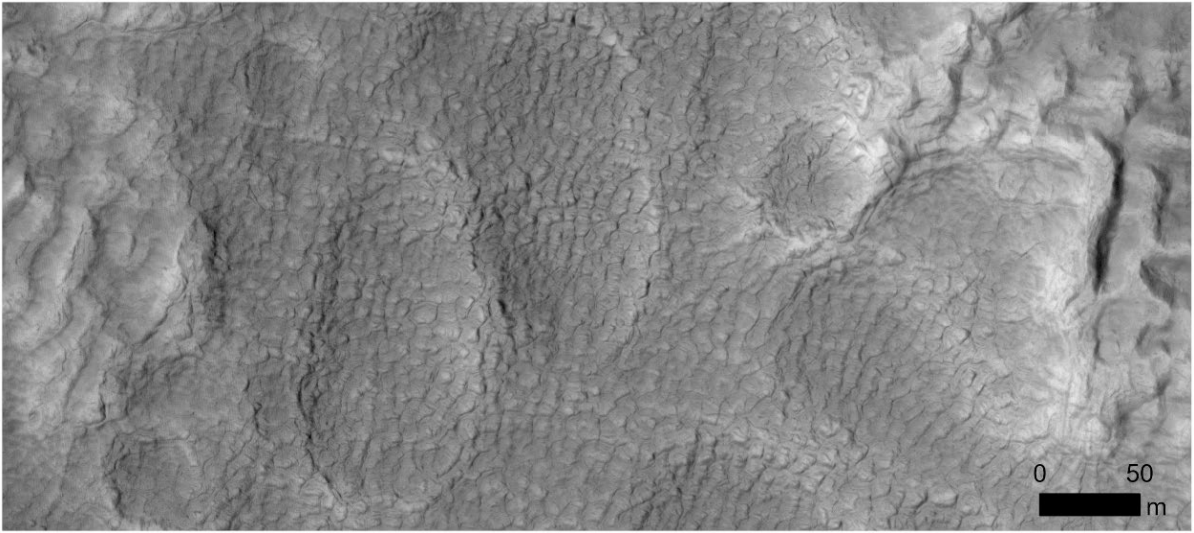

**Fig. 6:** Segment of light-toned terrain in which high-centred polygons and polygonised

depressions occur. North is up. *HiRISE* Image credits: *NASA/JPL*/University of Arizona.

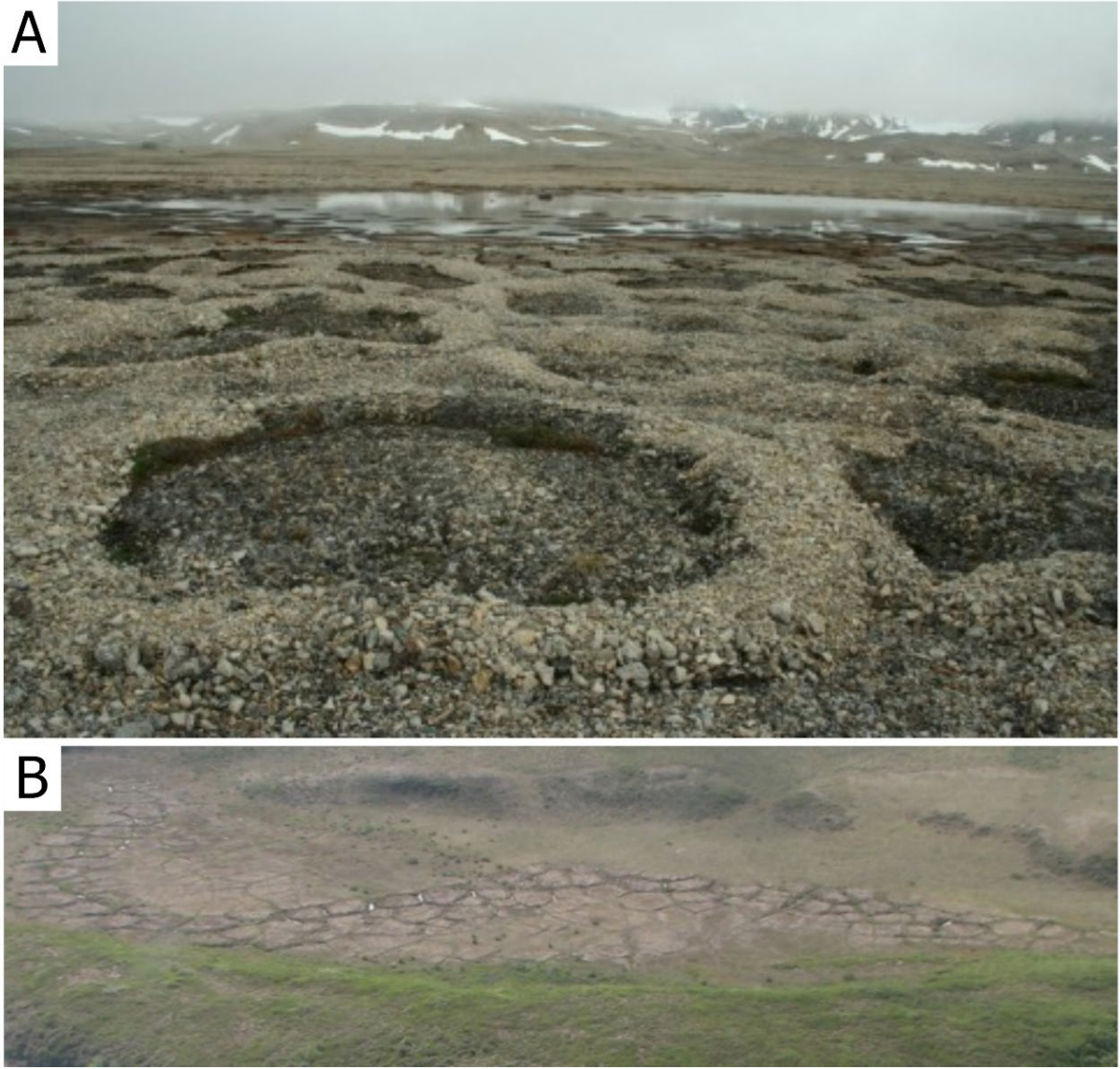

**Fig. 7: a)** Clastically-sorted circles, Kvadehukken, Svalbard. Photo credit and permission to reproduce granted: Ina Timling, Geophysical Institute, University of Alaska Fairbanks, 903 Koyukuk Drive, Fairbanks, Alaska, USA 99775). **b)** Oblique view of thermokarst-lake basin (alas) incised by polygons with centres slightly more elevated than the margins (Husky Lakes, midway between the coastal village of Tuktoyaktuk and Inuvik, on the eastern embankment of the Mackenzie River delta. Image credit: R. Soare.

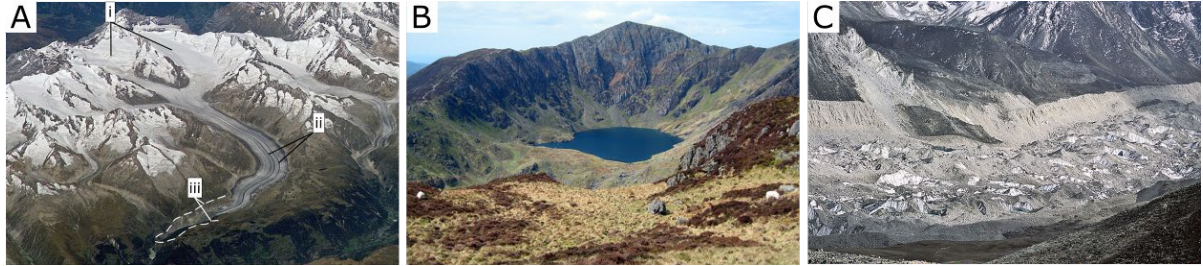

**Fig 8: a)** Grosser Aletschgletscher glacier, Switzerland from the International Space Station (Image ISS013-E-77377) looking *NNE*. Labelled are the source cirque **(i)**, medial moraines generated as debris from adjoining basins coalesces **(ii)**, and a latero-frontal moraine marking the terminus of the glacier **(iii)**. **b)** Cwm Cau, a cirque on the eastern face of Cadair Idris, Wales. **c)** A 'degraded' glacial surface on Khumbu glacier, Nepal. Spatial heterogeneity in debris thickness leads to high local ablation where debris is thinnest and the subsequent development of ice cliffs and meltwater ponding (e.g., Watson et al., 2017). Panel **b)** and **c)** are reproduced with permission from https://www.swisseduc.ch/glaciers/, photo credit: M.J. Hambrey.

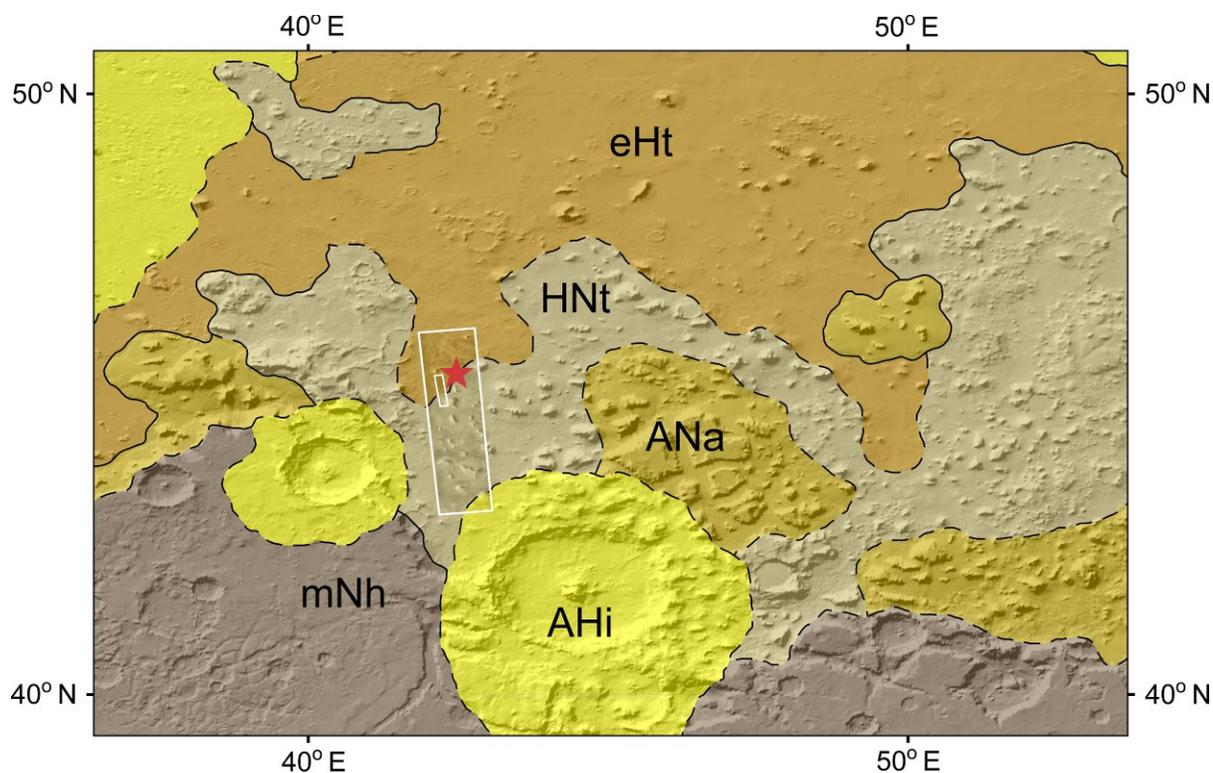

**Fig. 9:** Geologic units mapped by Tanaka et al (2014) centered on Protonilus Mensae and overlying *MOLA* shaded relief. Unit boundaries are marked with black lines, dashed where inferred. The larger white rectangle is the image boundary of *CTX* image F21_044083_2248_XI_44N317W; the smaller white rectangle is the image boundary of *HiRISE* image ESP_028457_2255. Red star marks the location of the topographical depression possibly exposing the basement of unit *eHt*. The relative elevation of the depression (see **Fig 11a**) is lower than the two other reference elevations for the dark and light-toned terrains at their geological contact immediately to the southwest **(Fig. 11a)**. Image credit *NASA/JPL*/Arizona State University.

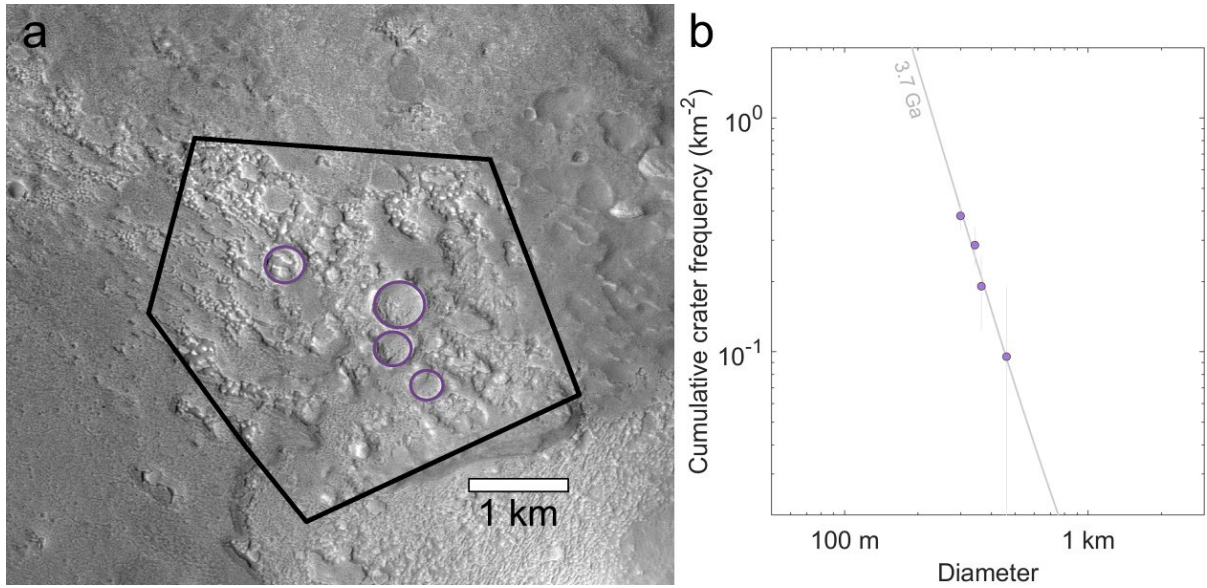

**Fig. 10: a)** Magnified *CTX* image of divot-like topographic depression (~10 km² area) referenced in **Fig. 9**. Four partially exposed craters $D \geq 300$ m (purple) are observed. **b)** The cumulative *CSFDs* for the craters compares well with a 3.7 Ga model isochron from Hartmann (2005) and with the age of the *eHt* unit identified by Tanaka et al., (2014).

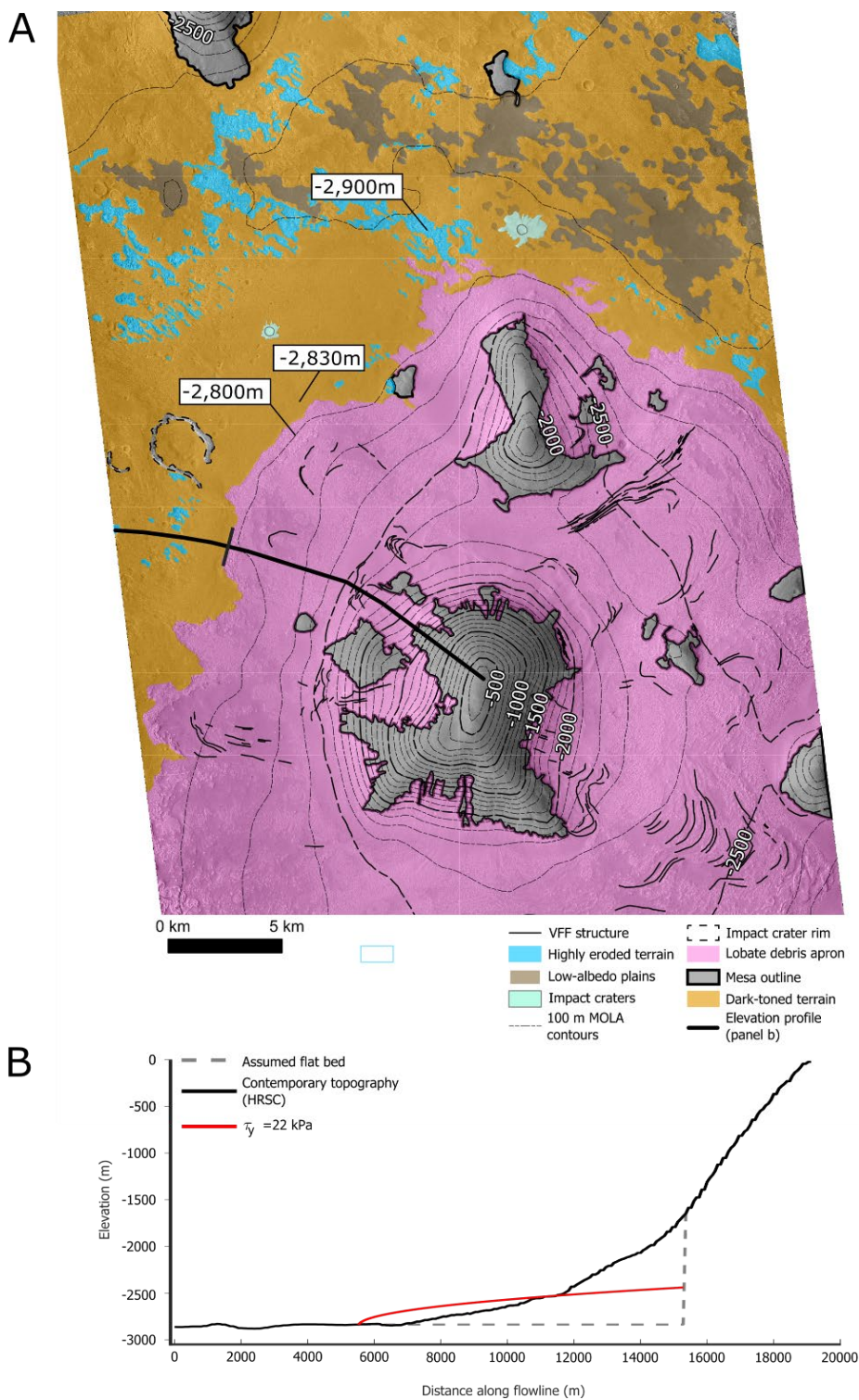

**Fig. 11: a)** Planimetric view and contour profile of units *HNt/eHt* based on *CTX* image. Contours

derived from the global *MOLA* elevation dataset (Zuber et al., 1992). North is up. Image credit: *NASA/JPL*/Arizona State University. **b)** Modelled and measured surface profiles for the *VFF* shown in **a)**. Red line derived from 2D model of perfect plasticity, black line derived from *MOLA* elevation along the profile shown in **a)**. The modelled profile is a poor fit for the measured profile of the *VFF* and a thicker ice mass extending beyond the geological contact is predicted based upon our initial assumptions.

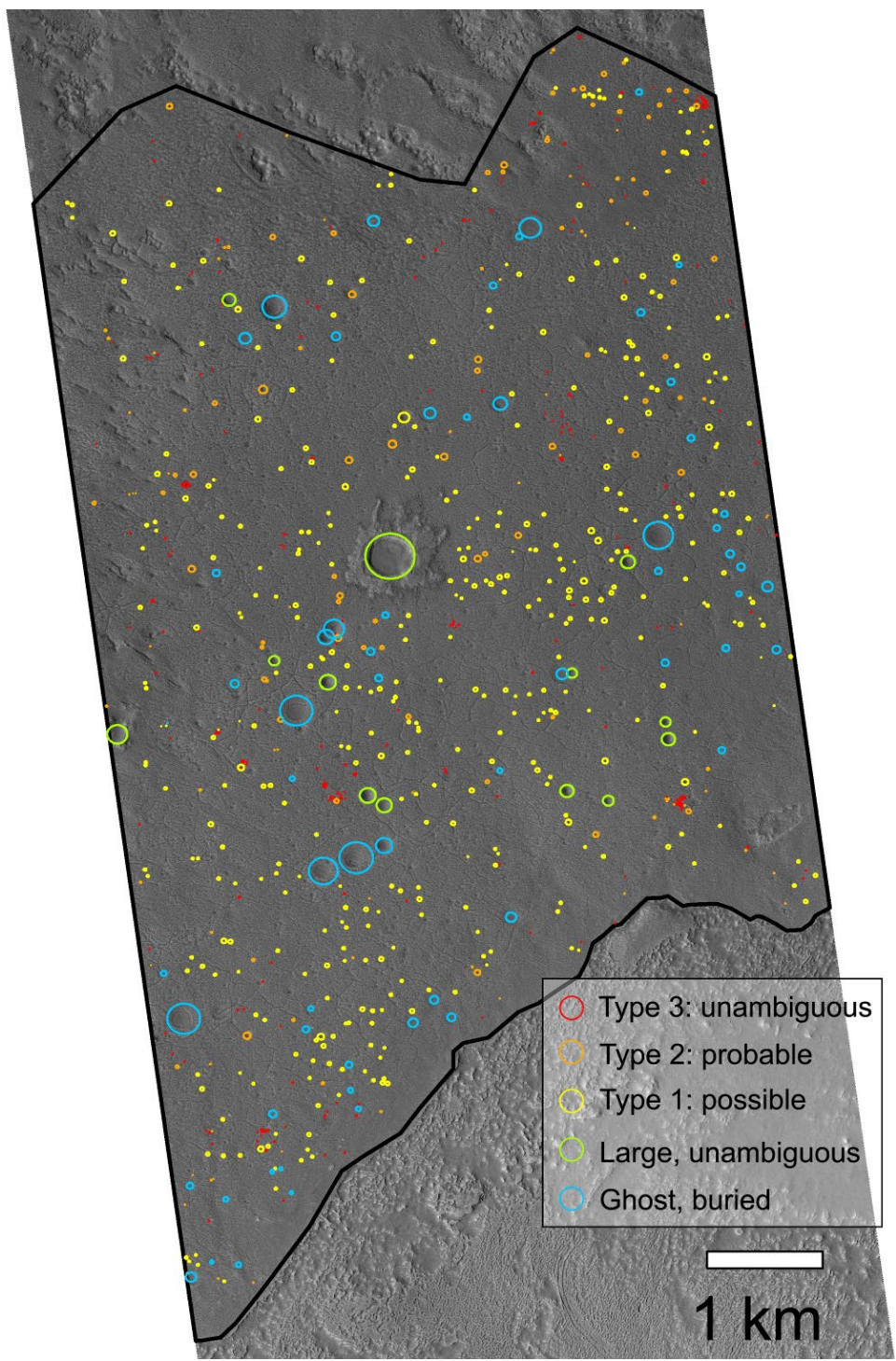

**Fig. 12:** Crater count area in (sample) segment of dark-toned terrain (outlined in black) (*HiRISE* image ESP_028457_2255). Crater colours indicate crater class: *red* -Type 3, *orange* - Type 2, *yellow* - Type 1, *green* - Large (*D* > 80 m), unfilled, *blue* - filled ghost/buried (see **Figs.**

**13-14**). Type 0 craters are excluded as their impact origin is highly uncertain. The Type 3, Type 2, and the large, unfilled craters represent a population of craters confidently identified as having accumulated on the surface of the lithic unit (*red*, *orange*, and *green* markers) and are used to generate the combined *CSFDs* in **Fig. 13**. Image credit: *NASA/JPL*/University of Arizona.

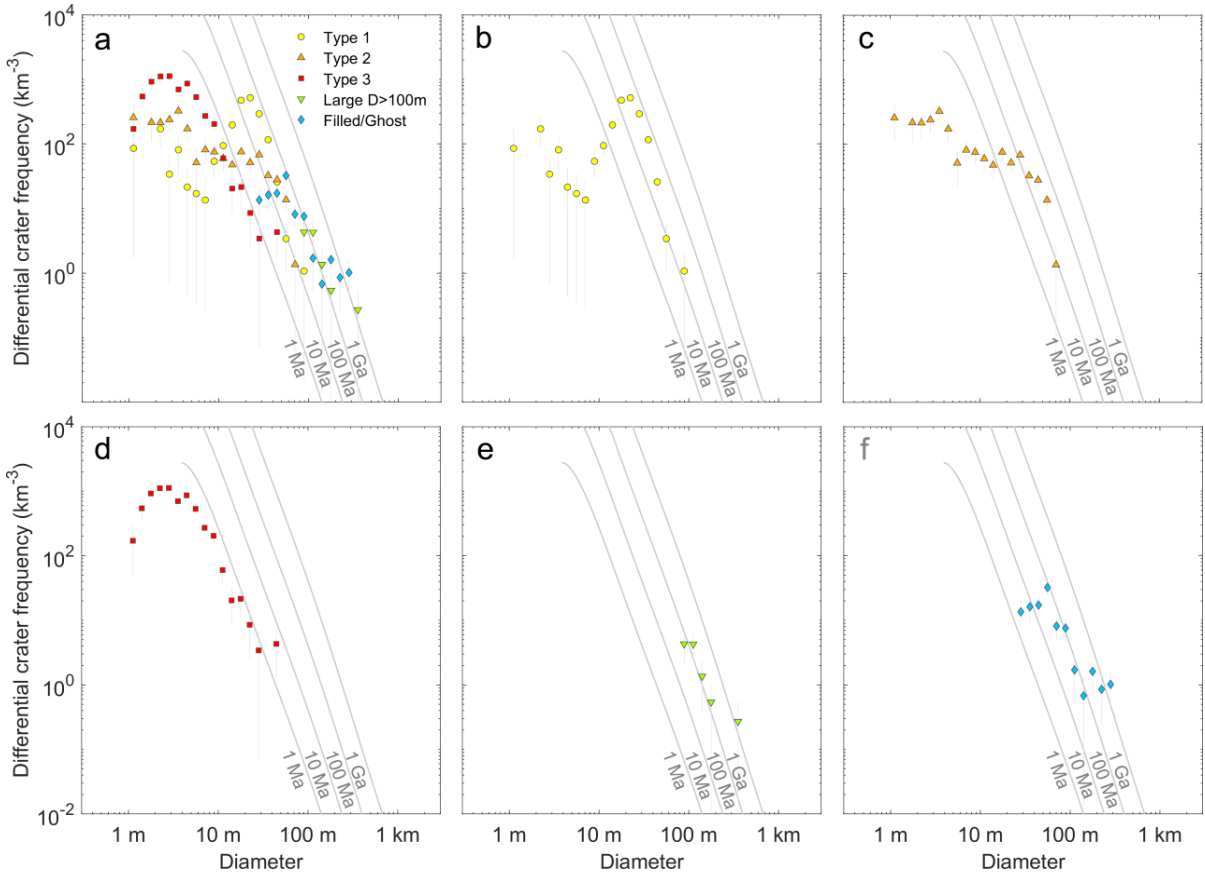

**Fig. 13 a)** Differential Crater Size-Frequency Distribution (*CSFD*) of each categorized crater class excluding *Type 0* because it is not deemed to be impact related. Marker colors correspond to colors of the mapped craters in **Fig. 12**. For clarity, **(b-f)** show the CSFDs individually: **b)** *Type 1*, **c)** *Type 2*, **d)** *Type 3*, **e)** *unfilled large D ≳ 80 m*, and **f)** *filled* (buried/ghost) craters. *Type 3* (*red*) craters represent smaller, fresh craters and suggest the upper surface has been stable for ~1 Myr against erosion and modification. *Type 2* craters are heavily modified but still visible after 10's of Myr. The larger unfilled craters ($D > 80$ m) suggest the material is at least ~100 Ma. The *CSFD* of the Type 1 craters *b)* peak at $D \sim 10$ - 40 m. Since the texture of the terrain occurs at this length-scale this could indicate that many of the features mapped as *Type 1* craters have been misidentified. Thus, this class of craters has been excluded in the combined isochrons in **Fig. 14**. Their exclusion makes little difference on the age interpretation. Model isochrons (gray) are for ~1 Ma, ~10 Ma, ~100 Ma, ~1 Ga for all figures.

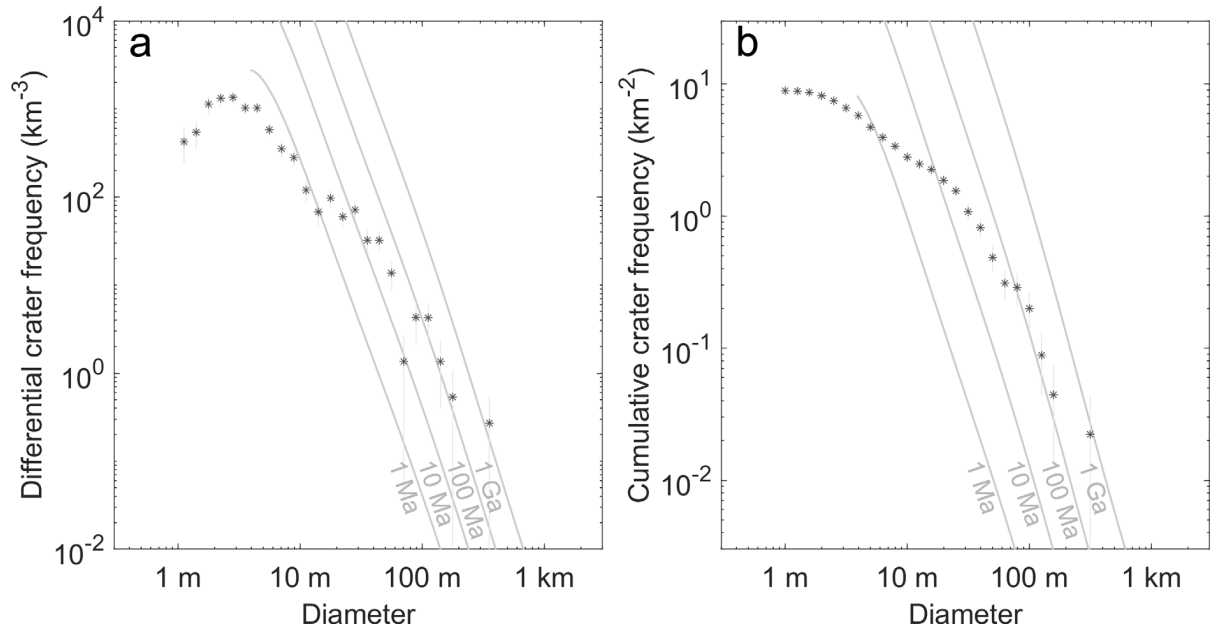

**Fig. 14: a)** Differential and **b)** cumulative *CSFDs* of the craters confidently interpreted as impact related in the dark-toned unit (*Types 2* and *3*, and the large unfilled craters from **Fig. 4a**. The largest craters (*D* > 80 m) suggest the dark-toned unit is >~100 Ma.

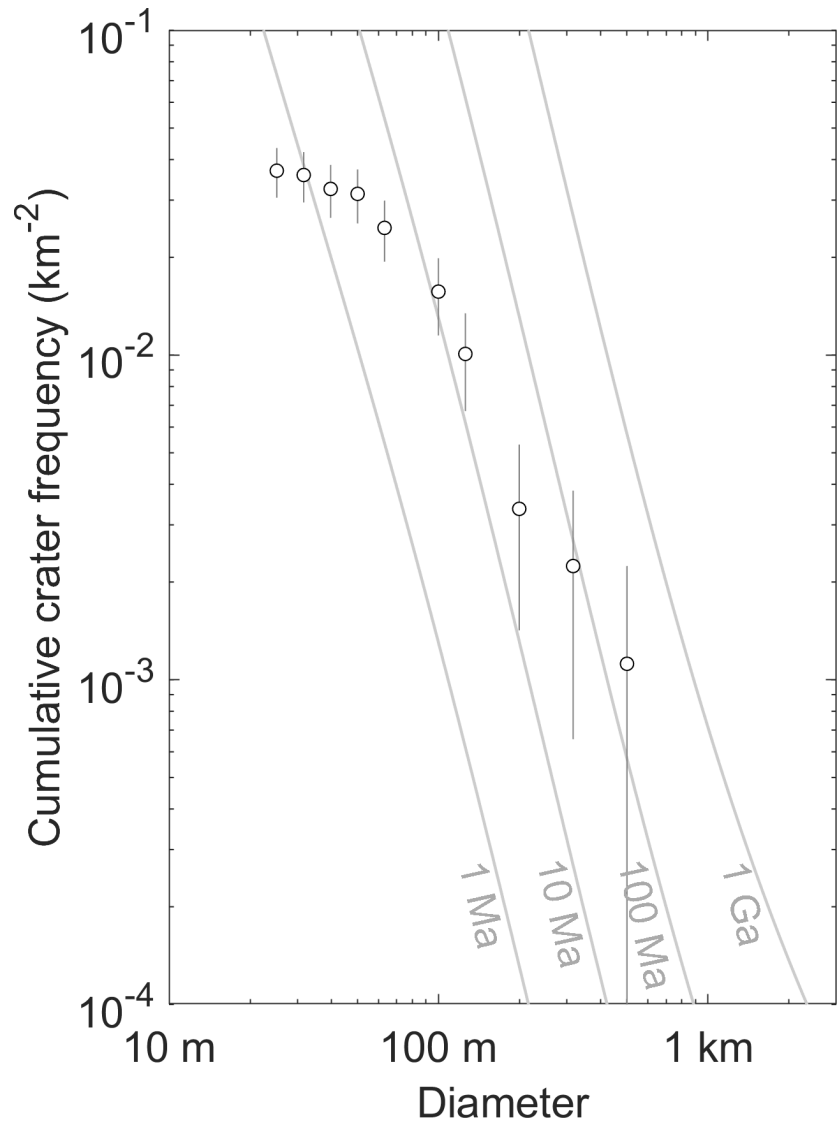

**Fig. 15:** Cumulative *CSFD* of the light-toned surface (area mapped as 'Lobate debris apron' in

**Fig. 11a)** representing an area that is 893 km². A total of 33 craters were identified as unambiguously impact related based on size and morphology suggesting a crater retention age ~10 - ~100 Ma for craters *D* > 100 m.